# Ultraviolet and Near-Infrared Dual Band Selective-Harvesting Transparent Luminescent Solar Concentrators

Chenchen Yang,[1,†] Wei Sheng,[2,†] Mehdi Moemeni,[2] Matthew Bates,[1] Christopher K. Herrera,[1] Babak Borhan,[2] Richard R. Lunt[1,3,*]

[1] Department of Chemical Engineering and Materials Science, Michigan State University, East Lansing, MI, 48824 USA

[2] Department of Chemistry, Michigan State University, East Lansing, MI, 48824 USA

[3] Department of Physics and Astronomy, Michigan State University, East Lansing, MI, 48824 USA

* Correspondence: rlunt@msu.edu

[†]The authors C. Y. and W. S. made equal contribution to the work.

## Abstract

Visibly transparent luminescent solar concentrators (TLSCs) can optimize both power production and visible transparency by selectively harvesting the invisible portion of the solar spectrum. Since the primary applications of TLSCs include building envelopes, greenhouses, automobiles, signage, and mobile electronics, maintaining aesthetics and functionalities is as important as achieving high power conversion efficiencies (*PCE*s) in practical deployment. In this work, we combine massive-downshifting phosphorescent nanoclusters and fluorescent organic molecules into a TLSC system as ultraviolet (UV) and near-infrared (NIR) selective-harvesting luminophores, respectively, demonstrating UV and NIR dual-band selective-harvesting TLSCs with *PCE* over 3%, average visible transmittance (*AVT*) exceeding 75% and color metrics suitable for the window industry. With distinct wavelength-selectivity and effective utilization of the invisible portion of the solar spectrum, this work reports the highest light utilization efficiency (*PCE* × *AVT*) of 2.6 for a TLSC system, the highest *PCE* of any transparent photovoltaic device with *AVT* greater than 70%, and outperforms the practical limit for non-wavelength-selective transparent photovoltaics.

## MAIN TEXT

## Introduction

Building-integrated photovoltaic technologies (BIPV) can convert new and existing surfaces into power-generating sources, which simultaneously enables the operation of autonomous landscapes and reduces distribution losses.[1–5] To maximize the output from the incident solar energy, integration can be advantageously deployed over the entire building envelope including the facades and rooftop areas. For the siding and window area this requires that the aesthetic quality is not compromised by the BIPVs. Visibly transparent photovoltaic (TPV) technologies typically aim to harvest the ultraviolet (UV) and near-infrared (NIR) portions of the incident solar spectrum and allow the visible (VIS) light to pass through, converting the invisible portion of light into electricity to supply on-site energy.[1,3,5] Additionally, TPVs can be readily integrated onto other smaller area applications including greenhouses, (electric) automobiles, (autonomous) mobile electronics and textiles while improving the energy utilization efficiency.[1,4]

Practical deployment of TPV technologies requires both high power conversion efficiency (*PCE*) and high aesthetic quality, including high average visible transmittance (*AVT*) and color rendering index (*CRI*).[1,29] Therefore, it is beneficial to maximize the light harvesting in the invisible portion of the solar spectrum and simultaneously fine-tune the absorption cut-off edges precisely at the UV/VIS and VIS/NIR borders to maximize the visible transmission (435 – 675 nm).[1,3,5] In the past 5 years, efforts have been made to achieve high *PCE* and visible transparency in TPVs. For example, the bandgaps of organometallic halide perovskite materials were sensitively tuned by compositional engineering for UV-selective-harvesting TPVs;[6,7] a series of novel low-bandgap polymer donors and non-fullerene acceptors have been applied in organic PV devices,[8,9] and excellent photovoltaic performance with distinct NIR selectivity has been demonstrated;[6,8,10–13] tandem architectures have also been utilized in TPVs to selectively harvest both UV and NIR portion of the incident solar spectrum, substantially reducing thermal losses and improving the output photovoltaic performance despite limitations imposed by current-voltage matching;[6,11] the utilization of optical outcoupling layers for VIS photons and various types of transparent electrodes can simultaneously enhance the visible transparency and the utilization of invisible photons.[10] Currently, the *PCE* of thin-film TPVs have reached ~8-10%, however, the highest reported *AVT* is around 40-50% due to considerable parasitic absorption from the electrodes, active layers, and non-ideal wavelength-selectivity.[10,11,14]

Alternatively, transparent luminescent solar concentrators (TLSCs) optically shift the solar energy conversion to edge-mounted traditional PV cells via waveguided photoluminescence (PL) via total internal reflection. The lack of electrodes over the active area enables TLSCs with wavelength-selectivity to achieve the highest possible visible transparency while improving defect tolerance and eliminating the need for electrode patterning. This can help circumvent several of the challenges for thin-film TPVs and simplify the manufacturing.[1,3,4] Much of the previous work on TLSCs with NIR harvesting have absorption profiles that have limited UV capture with *PCE*s up to around 1% and *AVT*s above 70% for a light utilization efficiency (*LUE*) of 0.7. Note that $LUE = PCE \times AVT$, and was introduced to enable a comparison between various TPV technologies against theoretical limits of varying *AVT*.[15] The highest reported and certified semitransparent LSC devices based on inorganic nanocrystals have reported a *PCE* of 2.2% with an *AVT* of 44% (*LUE* of 0.97) and a brown color.[16] Multiple luminophores with various wavelength-selectivity can be incorporated into the LSC waveguide to maximize the spectral coverage of light harvesting,[17–24] enhance photovoltaic performance,[2,25–27] and balance the color neutrality.[1,5,28,29] However, the coupling or reabsorption between different luminophores often leads to a reduction in the efficacy of this approach.[30,31] In this work, we simultaneously introduce highly luminescent phosphorescent nanoclusters (NCs) and fluorescent organic molecules into TLSCs as isolated UV and NIR selective-harvesting luminophores, respectively. The nanoclusters selectively harvest UV photons while exhibiting near-unity photoluminescence quantum yields (*QY*) and massive downshift of the luminescence into the NIR, without the use of heavy or toxic elements like lead.[41,45] To effectively pair these emitters and prevent parasitic reabsorption loss of the nanocluster emission in the NIR absorbing organic fluorophore we show a strategy to isolate the absorption/emission bands. The corresponding dual-band selective-harvesting TLSC exhibits *PCE* over 3% due to the effective utilization of the invisible photons and high *QY*s of the luminophores. Distinct UV and NIR selectivity offers the TLSC excellent aesthetic quality (*AVT* over 75% and *CRI* of 90). These down-shifting dual-band TLSCs also show good photostability with minimal degradation over more than 700 hours of continuous 1 Sun illumination. This work reports TLSC devices with the highest *PCE* at the highest transparency, the highest *LUE*, and demonstrates a novel design to effectively utilize the solar spectrum in a highly aesthetical approach.

**Results**

The dual-band TLSC device is composed of two distinct waveguides as shown in **Figure 1**A with the UV component coated in polymer matrix on one waveguide and the NIR component on the other. An air gap is utilized to optically isolate the waveguided luminescence in each panel to prevent parasitic reabsorption and retain scalability. For more practical deployment, this air gap can be replaced with a low-index polymer,[32–34] metal oxide,[35–39] or glue with little change in the performance.[40]

The top UV component is based on phosphorescent hexanuclear nanoclusters, where the chemical structure of $Cs_2Mo_6I_8(CF_3CF_2COO)_6$ NC is shown in Figure 1B (the synthesis is described in the Experimental Section). Substitution of the apical halide positions has been shown previously to be an effective approach to increase *QY*s above 50%.[41–46] Both X-ray diffraction (XRD)[42] and mass spectrometry[41,43] are used to confirm the formation of the synthesized products (See Supplemental Information Note 1 and 2 for detail). Various terminating ligands ($(CF_3)_n$ chain length) were synthesized and tested to maximize the *QY* with the composition above providing the highest value. We note that the chemical composition of the NC does not contain any hazardous heavy metal ions, which makes the deployment more environment-friendly. The TLSC waveguide was made by drop-casting NC/polymer mixture onto square borosilicate glass sheets to form uniform composite films. The normalized absorption and emission spectra of the NC in polymer are shown in Figure 1B. The spectra show band absorption cut-off at the UV/VIS border and NIR emission onset at the VIS/NIR border with a massive downshift over 300 nm and a corresponding *QY* of 80±5 in polymer matrix (75±5% in acetonitrile), which makes these NCs a good UV selective-harvesting luminophore for TLSC applications.[46]

The bottom waveguide is based on fluorescent organic small molecules. In organic and molecular semiconductors light absorption originates from the transition from the ground state to excited molecular orbitals. The energy difference between the excited molecular states forms discontinuities in the density of states. Therefore, these energy gaps can be tuned to transmit visible photons in TPV applications. In this work, two different organic luminophores are demonstrated as NIR selective-harvesters: COi8DFIC (also referred to as O6T-4F),[6,8,9,13] which has been developed as a non-fullerene acceptor in organic photovoltaics with excellent performance; and a BODIPY derivative with high *QY* in the NIR (details of the syntheses are provided in the Experimental Section).[35] The molecular structures, normalized absorption and emission spectra of these NIR components in polymer matrix are shown in Figure 1C and D, respectively. Similarly, the NIR selective-harvesting waveguide was also made by drop-casting dye/polymer mixture onto glass sheets to form a uniform composite film. The absorption peak of COi8DFIC is at 745 nm and the emission peak is at 808 nm, resulting in a Stokes shift of ~60 nm and *QY* of 25±3% in polymer matrix (23±1% in chlorobenzene). Compared to COi8DFIC, the absorption peak of BODIPY is narrower with a smaller Stokes shift (10 nm), but the significantly higher *QY* of 40±3% in polymer matrix (41±2% in hexane) is among the highest values for this NIR emission range. Moreover, we have shown previously that Stokes shift is not always well correlated to performance and a more important parameter to analyze in the modified overlap integral (*OI*: 0.015, 0.40 and 0.56 for UV component with NC, NIR component with COi8DIFC and NIR component with BODIPY, respectively.),[15] which indicates a similar level of reabsorption probability between the two emitters.[15,47]

For optimizing LSCs it is advantageous to select an edge-mounted PV cell with a bandgap bordering the emission edge of the luminophores. This allows all the waveguided PL to be collected and converted to electricity while minimizing the voltage losses due to thermalization. Thus, the voltage of the LSC system is increased. With all three emission edges of NC, CO$_i$8DFIC and BODIPY below 900 nm as shown in Figure 1B-D, GaAs is a nearly ideal edge-mounted PV cell choice for these luminophores to maximize the overall photovoltaic performance. GaAs cells are mounted on two edges for current density versus voltage (*J-V*) measurements and on one edge of the dual waveguide for external quantum efficiency ($EQE_{LSC}$) measurements following the standardized procedures outline elsewhere, where both are accordingly corrected for the equivalent four-edge mounting (See Experimental Section for detail). [28]

Single-band TLSC devices with one lumiphore were first fabricated and optimized based on concentration. Dual-band TLSC devices with two luminophore combinations (NC+CO$_i$8DFIC and NC+BODIPY) were then fabricated and their photovoltaic performance was characterized. For comparison, the TLSC with NC-only (10 mgmL$^{-1}$) was added as a reference device. The *J-V* characteristics of these TLSCs (active area of 5.08×5.08 cm$^2$ and total waveguide thickness of 0.635 cm) measured under AM 1.5G illumination are shown in **Figure 2**A. When a PV cell is edge-mounted onto an LSC waveguide, the LSC-PV system should be treated as an integrated photovoltaic device, and the input solar photon flux is received by the area of the front surface of the LSC waveguide ($A_{LSC}$) rather than the area of the edge-mounted PV ($A_{Edge}$), just as with any PV system. The NC-only TLSC shows short-circuit current density ($J_{SC}$) of 2.5±0.2 mAcm$^{-2}$, open-circuit voltage ($V_{OC}$) of 1.01±0.01 V and fill factor (*FF*) of 80±1%, resulting in a *PCE* of 2.0±0.1%. As the organic molecules are added into TLSCs with the same UV component (NC concentration is kept at 10 mgmL$^{-1}$), the $J_{SC}$ values are improved to 3.6±0.2 mAcm$^{-2}$ and 3.8±0.1 mAcm$^{-2}$ while exhibiting similar $V_{OC}$ and *FF*. This results in corresponding *PCE*s that reach 2.9±0.1% and 3.01±0.07% for NC+CO$_i$8DFIC and NC+BODIPY TLSCs, respectively, with color metrics suitable for the window industry. A champion device *PCE* of 3.65% is reached with higher NC concentration, however, the corresponding color metrics are outside the range of suitable for the window industry (See Supplemental Information Note 6 for detail). The average position-dependent $EQE_{LSC}(\lambda)$ spectra are shown in Figure 2B. For NC-only TLSC the $EQE_{LSC}$ contribution originates only from the light absorption of the UV selective-harvesting NC. Neither Rayleigh scattering (caused by particle aggregation) nor direct illumination of the edge-mounted PV is observed from the $EQE_{LSC}$ profile. This also indicates that the haze from the devices is negligible, which is confirmed with optical spectroscopy. For the NC+COi8DFIC and NC+BODIPY TLSCs, both UV and NIR peaks appear in their corresponding $EQE_{LSC}$ spectra, which result from the dual-band selective-harvesting. The $EQE_{LSC}$ peak positions match the absorption spectra of the corresponding luminophores. The $EQE_{LSC}$ peak heights are constrained by both the luminophore *QY* values and the absolute absorption spectra. With the same NC concentration, the UV contribution is nearly the same for all three TLSCs. Both slightly higher absolute absorption peak and significantly higher *QY* of the BODIPY results in a substantially higher $EQE_{LSC}$ peak compared to that of the COi8DFIC for this device size. As one of the most important consistency checks for any photovoltaic device, the $J_{SC}$ values extracted from *J-V* characteristics are confirmed by the integrated $J_{SC}$ ($J_{SC}^{Int}$) from $EQE_{LSC}(\lambda)$. The $J_{SC}^{Int}$ values are 2.42 mAcm$^{-2}$, 3.60 mAcm$^{-2}$ and 3.89 mAcm$^{-2}$ for NC-only, NC+COi8DFIC and NC+BODIPY TLSCs, respectively, and match well with the $J_{SC}$ from the *J-V* curves. Although the $EQE_{LSC}$ peak of BODIPY in the NC+BODIPY TLSC is higher than that of the NC+COi8DFIC TLSC, the broad absorption width of COi8DFIC compensates for the lower absorption peak and *QY*, resulting in similar contributions from the NIR components but different aesthetic quality.

The series of position-dependent $EQE_{LSC}$ spectra can be used to understand the scalability of LSC systems. The dual-band TLSC systems were fabricated with larger dimension (active area of 10.16×10.16 $cm^2$), and the series of $EQE_{LSC}$ at various $d$ are plotted in Figure 2C and D for NC+$CO_i$8DFIC and NC+BODIPY TLSCs, respectively, where $d$ is the distance between the incident excitation beam and the edge-mounted PV cell along the centerline of the square waveguide (See Experimental Section for detail, and the corresponding photovoltaic performance are tabulated in Supplemental Information Note 6).[28] Both UV and NIR peak values of each individual scan were extracted, normalized and plotted as a function of $d$ in Figure 2E and F. The NC-only, $CO_i$8DFIC-only and BODIPY-only TLSCs were also fabricated as references (See Supplemental Information Note 3 and Note 6 for detail). With the massive downshift of the NC, the reabsorption loss is so negligible that the $EQE_{LSC}$ peak values in the UV of the NC-only, NC+$CO_i$8DFIC and NC+BODIPY TLSCs stay nearly constant as $d$ increases. With the absorption and emission profiles of all the luminophores as inputs, optical simulations are given in Supplemental Note 7, which shows that the UV component is suitable for scaling to the practical size over 1 m and the NIR components are suitable for applications around 0.3 m that could be further improved with Stoke shift engineering principles shown previously.[47] However, due to significantly stronger overlap between the absorption and emission spectra for both $CO_i$8DFIC and BODIPY, the reabsorption loss leads to a more pronounced decay of the NIR peak values compared to the UV peaks. As shown in Figure 2E and F, the UV and NIR peak decay behaviors of the dual-band TLSCs strongly resemble those of the NC-only, $CO_i8DFIC$-only and BODIPY-only TLSCs, respectively. Given the similarity in the decay trend in each range of the $EQE_{LSC}$ spectra, the isolation of the waveguides effectively enables total internal reflection within each waveguide so that the UV and NIR components operate nearly independently. Improvements in scalability to the largest device sizes are likely achievable via Stoke shift engineering which has led to values over 100 nm for single-fluorescent emitters (See Supplemental Information Note 7 for detail), and more specifically, following chemical approaches that reduce overlap integrals.[47,48]

Aesthetic quality is equally important as photovoltaic performance for any TPV device, which determines whether a TPV device can be deployed in certain practical applications.[1,3,5,28,29] The transmittance spectra ($T(\lambda)$) of the NC-only, NC+$CO_i8DFIC$ and NC+BODIPY TLSCs are plotted in **Figure 3**A along with the photopic response of the human eye ($V(\lambda)$) for comparison. NC shows an absorption cut-off edge at the UV/VIS border and BODIPY exhibits a NIR-band absorption onset at the VIS/NIR border. However, the broad NIR-band absorption of $CO_i8DFIC$ extends into the red/NIR range, leading to lower visible transmittance with a slight blue tint. $T(\lambda)$ is used to quantify the main figures of merit for aesthetic quality: *AVT*, *CRI* and CIELAB color space coordinates ($a^*$, $b^*$). All three parameters are prominently utilized metrics in the window industry to assess overall transparency and color quality of glazing systems. With good UV selectivity, the NC-only TLSC shows *AVT* of 81.9% and *CRI* of 91.3. For the dual-band TLSCs, the *AVT* and *CRI* of the NC+$CO_i8DFIC$ TLSC drop to 65.6% and 82.9. With the better NIR selectivity of BODIPY, the *AVT* and *CRI* of the NC+BODIPY TLSC is improved to 75.8% and 88.3, respectively. Even more important are the color coordinates, which are discussed below in detail.

The photon balance is a necessary consistency check to confirm the validity of independent measurements including $EQE_{LSC}(\lambda)$, $T(\lambda)$ and $R(\lambda)$ at every wavelength ($EQE_{LSC}(\lambda) + T(\lambda) + R(\lambda) \leq 1$). The photon balance for all the TLSC devices in this work is shown to be consistent in **Figure 4**.[1,28,29]

**Discussion**

Because color coordinates of glazing systems are often utilized as a strict criteria for product viability in the window industry, the impact of NC concentration on aesthetic quality and photovoltaic performance of the NC-only, NC+$CO_i$8DFIC and NC+BODIPY TLSCs is systematically studied for both performance and aesthetics. The CIELAB color space coordinates ($a^*$, $b^*$) are commonly utilized to assess acceptable ranges of color tinting for products in the glass and glazing industries ($-15 < a^* < 1$ and $-15 < b^* < 15$ for many mass market architectural glass products). As the "reference light source" for TPVs, the incident AM 1.5G is at the origin (0, 0) (as colorless or neutral),[1,29] and the ($a^*$, $b^*$) coordinates are plotted in Figure 3B as a function of NC concentration. These TLSCs are categorized into three groups: NC-only group, NC+$CO_i$8DFIC group and NC+BODIPY group, within each group NC concentration (1, 2, 5, 10 and 20 $mgmL^{-1}$) is the only variable. Additionally, the $CO_i$8DFIC-only and BODIPY-only TLSCs are included as references. As shown in Figure 3B, the incorporation of $CO_i$8DFIC or BODIPY leads to negative values of *a** due to the tail NIR absorption into red range. The *b** of NC-only, NC+CO$_i$8DFIC and NC+BODIPY TLSCs moderately increases as NC concentration increases from 1 to 5 mg $mL^{-1}$, while further increasing the concentration above 10 mg $mL^{-1}$ causes a dramatic drop in TLSC aesthetic quality and *b** values that are less acceptable to the window industry (*b** > 15).

Visibly absorbing semiconductor materials can also be utilized as active layers in TPV applications. Active layers with thin enough thickness or micro-segmented structure permits the transmission of a portion of visible light, which creates partial visible transparency.[49] However, there is a direct trade-off between photovoltaic performance and visible transmission in this approach. As shown in Figure 3B, any non-neutral absorption profile within 435-675 nm range can result in sharp drops in *AVT*, *CRI*, and increased deviation of (*a**, *b**) from the CIELAB origin. Therefore, this type of device is sometimes referred to as "semitransparent" PV or non-wavelength-selective TPV. Although the theoretical Shockley-Queisser (SQ) limit for an opaque PV is 33.1%, the *PCE* of a non-wavelength-selective TPV approaches 0% as the *AVT* approaches 100% - in the practical limit these devices approach 0% at *AVT*s around 85-90% due to reflections of double-pane encapsulation.[1,3,5] The SQ and practical *PCE* limit lines for non-wavelength-selective TPVs are shown in **Figure 5**A. For wavelength-selective TPVs or TLSCs which harvest only UV (< 435 nm) and NIR photons (> 675 nm), the corresponding SQ *PCE* limit is 20.6% with an *AVT* > 99%. The light-green shaded region reflect the practically achievable *PCE* and *AVT* combination with the wavelength-selective approach only, and the dark shaded green regions indicates the theoretical *PCE* and *AVT* combination with the wavelength-selective approach only. The *PCE* values as a function of *AVT* (60-100% range) of all three groups of TLSCs (including CO$_i$8DFIC-only and BODIPY-only references) are plotted in Figure 5A. Among all these devices, BODOIPY-only, NC-only, and NC+BODIPY (with 5 and 10 mgL$^{-1}$ NC concentrations) TLSCs are all above the practical *PCE* limit line for non-wavelength-selective TPVs for the first time due to the good NIR selectivity and high *QY*s. As the NC concentration increases from 1 to 10 mgmL$^{-1}$, the *PCE* vs. *AVT* trend line of the NC-only group maintains a trend nearly parallel to the practical SQ *PCE* limit line until it starts to deviate above 10 mgmL$^{-1}$ due to tail absorption extending into the visible range. The NC+CO$_i$8DFIC group and NC+BODIPY groups also show a similar trend as NC concentration increases, and the incorporation of the NIR component significantly improves the *PCE* of the dual-band selective-harvesting TLSC system over 3% (up to 3.65% with 20 mgmL$^{-1}$ NC concentration) with modestly reduced *AVT*.

Light utilization efficiency provides a metric for systematically comparing TPVs with different levels of *AVT* values on the same scale. *LUE* of all the TLSCs as a function of their corresponding *AVT* along with the SQ and practical *LUE* limit lines are plotted in Figure 5B. Although both the air gap and the tail of the NIR absorption into red range leads to a slightly reduced *AVT* level, the *LUE* still gains significant improvement stemming from the dual-band selective-harvesting. Literature reports are included as background in both *PCE* vs. *AVT* and *LUE* vs. *AVT* plots for comparison (See Supplemental Information Note 6 for detail). Among all the TLSCs, the NC+BODIPY shows the best *LUE* of 2.61 at an *AVT* of 71.6%, the highest *LUE* value reported for a TLSC system by over a factor of 2.[16] However, it is a balanced combination of *PCE*, *AVT*, and *CRI* or (*a**, *b**) that is important to consider when choosing optimal and deployable devices. As shown in **Table**

**S1** the NC+BODIPY with 5 $mgmL^{-1}$ NC concentration is expected to be the most suitable TLSC device for real-world deployment as $b^*$ is < 15. We also note that the aesthetics of a TLSC depends on its $T(\lambda)$, reflectance ($R(\lambda)$) and emission ($PL(\lambda)$) spectra. $T(\lambda)$ determines the aesthetic quality observed from the transmitted side of the device as discussed above; whereas $R(\lambda)$ affects the aesthetic quality observed from the incident side, which can also be quantitatively evaluated using *CRI* and ($a^*$, $b^*$) based calculations. Since $T(\lambda) + R(\lambda) + A(\lambda) = 1$, where $A(\lambda)$ is the absorption spectrum of the TPV device, distinct UV and NIR wavelength-selectivity with a neutral absorption profile in VIS can lead to good color rendering observed from both sides for a TPV device.[29] However, due to the working principle of LSC devices, a portion of the photoluminescence (~25%) can escape from the top and bottom of the waveguide (via the escape cone), which can be observed as "glow" if the PL (or a portion of PL) resides in the VIS range. Such glow can also affect the aesthetics of a TLSC device and create an effectively colorful haze under illumination. In our case, we have designed all the emitters to effectively emit outside of the visible range. While there is a slight advantage in being able to recapture NIR emitted light from the escape cone of the top waveguide with the NIR absorber in the bottom waveguide, this effect is relatively small.

Looking ahead we consider strategies for further increasing the performance to approach the TPV and TLSC limit. The total photon flux at wavelengths < 435 nm is only ~8% of the AM 1.5G. Harvesting light at wavelengths > 435 nm can rapidly cause a yellowish or brown tint (large positive $b^*$ values), which are unacceptable for the majority of window applications. In contrast, the NIR range between 675 nm and the absorption cut-off of the edge-mounted PV cell (e.g. Si, GaAs, etc.) coincides with the peak of AM 1.5G photon flux, which has significantly more potential for power generation. Even with absorption extending into the red range, a resulting blue tint (negative $a^*$ value) is more visually acceptable, which offers more design freedom for NIR selective-harvesting luminophores and can even help to compensate poor $b^*$ values from yellow tinting. The *QY*s of various UV-absorbing luminophores including quantum-dots and nanoclusters have been gradually improved to more than 80% in recent years, and further improvement will likely be rather limited.[6,21,22,24,36,38,39,43,46,47,50] By comparison, the *QY* of NIR luminophores currently ranges from 20-40%, including the compounds demonstrated in this work. However, there is still promise via chemical design to improve the *QY* closer to 60-80%. Improving the *QY* of NIR luminophores can effectively lead to performance improvement without changing the optical properties. This is reflected in Figure 2B for the NC+BODIPY TLSC: although the NC peak is much stronger than the BODIPY peak (due to higher *QY* and less reabsorption loss of the NC), the contribution from the NIR component is comparable to that from the UV component (2.4 $mAcm^{-2}$ from NC vs. 1.5 $mAcm^{-2}$ from BODIPY). Thus, future improvements in TLSCs can result from: 1) the improvement of the *QY* of the NIR selective-harvesting luminophores (allowing 2-3 times of enhancement in the NIR contribution without changing the aesthetics); 2) sharper wavelength-selectivity near the UV/VIS and VIS/NIR borders for higher visible transmittance and better color rendering; 3) separation of the absorption and emission spectra of the NIR luminophores to suppress the reabsorption loss. Considering a dual-band TLSC with *QY*s of ~80% in both UV and NIR components and nearly ideal wavelength-selectivity, the overall *PCE* would be ~7% with both $AVT > 80\%$ and $CRI > 90$. This *PCE* and *AVT* combination is well above the practical and theoretical SQ *PCE* and *LUE* limit lines shown in Figure 4A and B, and would be suitable for deployment in most practical applications.[1]

Long lifetime performance is another key feature in real-world deployment. The photostability of all the NC-only, NC+$CO_i$8DFIC and NC+BODIPY TLSCs were studied and are shown in **Figure 6** A-C. Three key parameters were chosen to evaluate the photostability of the TLSC devices, and these parameters were normalized by the corresponding initial values: $A(\lambda)$ spectrum is used to monitor the degradation of total light absorption for each luminophore; $EQE_{LSC}(\lambda)$ spectrum can be used to represent the degradation of the contribution of each luminophore to the overall photovoltaic performance; internal quantum efficiency ($IQE_{LSC}(\lambda) = EQE_{LSC}(\lambda) / A(\lambda)$) is the $EQE_{LSC}(\lambda)$ value normalized by the $A(\lambda)$ at each wavelength, is used to analyze the photoluminescence stability of each luminophore under constant illumination of 1 Sun. All three parameters of the NC-only and the UV components of both dual-band TLSCs remain nearly constant after 700 hours of constant illumination. With the UV component as the top waveguide UV photons are filtered, which helps to minimize any degradation of the organic luminophores in the bottom NIR component.

Finally, we note that the use of two waveguides with edge-mounted PVs in a TLSC can add manufacturing complexity, however, it is analogous to the deployment of double-pane windows (insulating glazing) with a low-E coating so that a dual-waveguide TLSC will likely only lead to a small incremental cost on a premium window.

In summary, by combining highly emissive NIR phosphorescent hexanuclear metal halide nanoclusters and NIR organic luminophores as isolated UV and NIR selective-harvesting luminophores, respectively, we have designed and demonstrated dual-band selective-harvesting TLSC devices. Harvesting invisible photons from both UV and NIR portions of solar spectrum leads to *PCE* > 3%, with good wavelength-selectivity that results in *AVT* > 75% and *CRI* of 90 (*LUE* > 2.5). This approach could lead to devices with efficiency approaching 10% as NIR *QY*s are further improved. This work demonstrates the potential of TLSCs to be deployed as power-generating sources in multiple applications with high photovoltaic performance, excellent aesthetic quality, and long-term photostability. With simple and low-cost manufacturing, this technology is able to offer a promising approach to utilize solar energy in entirely new ways.

## Materials and Methods

*Nanocluster Synthesis*:

1) $Cs_2Mo_6I_{14}$: $MoI_2$ powder (2A Biotech) was uniformly mixed with CsI powder (Sigma-Aldrich) with a stoichiometric ratio of 3:1. The mixture was then transferred into a quartz ampule (12 cm long, 1.5 cm diameter), and the ampule was sealed under vacuum. The ampule was heated at the reaction temperature of 750 °C for 72 hours to form $Cs_2Mo_6I_{14}$. After cooling down to room temperature, the powder in the ampule was dissolved in acetone (wine-colored solution) and the undissolved impurity (unreacted black powder) was filtered out. The acetone was removed by rotary evaporation to obtain red $Cs_2Mo_6I_{14}$ powder. Powder XRD pattern of $Cs_2Mo_6I_{14}$ was collected to confirm the product (See Supplemental Information Note 1 for detail).

2) $Cs_2Mo_6I_8(CF_3CF_2COO)_6$: $Cs_2Mo_6I_{14}$ was weighed and dissolved in acetone in a flask, and silver pentafluoropropionate ($CF_3CF_2COOAg$) (Sigma-Aldrich) was added into the $Cs_2Mo_6I_{14}$ solution with a stoichiometric ratio of 6:1. The reaction was kept in the dark in a nitrogen atmosphere for 48 hours. After the ligand exchange reaction, the precipitated AgI was filtered out and the solution (cider-colored) was dried by rotary evaporation to obtain orange $Cs_2Mo_6I_8(CF_3CF_2COO)_6$ powder. The powder was purified by silica column chromatography (20% ethanal/80% acetone, gradually increasing to 100% ethanol) to yield the pure nanocluster product. Column chromatography was performed using Silicycle 60 Å, 35-75 μm silica gel. The final purification step boosts the NC *QY* by ~10% by eliminating the non-radiative impurities from the reactions. $Cs_2Mo_6I_8(CF_3COO)_6$ and $Cs_2Mo_6I_8(CF_3CF_2CF_2COO)_6$ nanoclusters were prepared by reacting $Cs_2Mo_6I_{14}$ and silver trifluoroacetate ($CF_3COOAg$) or silver heptafluorobutyrate ($CF_3CF_2CF_2COOAg$ ) with similar procedure. All the NC products were confirmed by high resolution mass spectrometry (Xevo G2-QTOF) (See Supplemental Information Note 2 for detail).

3) $CO_i8DFIC$[8,13] and BODIPY[35] syntheses follow the reported procedures from literature but are briefly summarized below (See Supplemental Information Note 8 for detail). Starting from lithiation of commercially available 3-bromothieno[3,2-b]thiophene **1**, followed by carbonylation and esterification was afforded intermediate **2** as a mixture of regioisomers. The ratio of desired isomer **2a** was enriched by recrystallization following our previous report.[15] The obtained material was directly subjected to subsequent Stille coupling, $BBr_3$ demethylation, lactonization, Grignard reaction, and Vilsmeier-Haack formylation, to furnish key precursor **4**. **$CO_i$8DFIC** was afforded by a final condensation with difluoroindanone **5**. The synthesis of **BODIPY** commenced with 2,3-dihydroxynaphthalene, which was converted to the corresponding dihydrazine **6**, and followed by formation of dihydrazone **7** for subsequent acid-catalyzed Fischer indole synthesis and decarboxylation to furnish the key building block naphthobipyrrole **8**. The BODIPY scaffold was then constructed by orthoformation in the presence of $POCl_3$ and following treatment with $BF_3{\cdot}OEt_2$.

*Module Fabrication*:

1) UV waveguide: $Cs_2Mo_6I_8(CF_3CF_2COO)_6$ nanocluster powder was weighed and dissolved in ethanol to prepare the solution at the target concentration (1, 2, 5, 10 and 20 $mgmL^{-1}$). The ethanol solution was mixed with mounting medium (Fluoroshield F6182, Sigma-Aldrich) at a volume ratio of 1:2.

2) NIR waveguide: $CO_i8DFIC$ or BODIPY was dissolved in dichloromethane to prepare the solution (100 $mgL^{-1}$ for BODIPY and 125 $mgL^{-1}$ for $CO_i8DFIC$). The dichloromethane solution was mixed with mounting medium (Shandon, Thermo Fisher Scientific) at a volume ratio of 1:1.

This mixture was drop-cast on 50.8 mm × 50.8 mm × 3.175 mm (for *J-V* characterization and averaged $EQE_{LSC}$ measurement) and 101.6 mm × 101.6 mm × 3.175 mm (for position-dependent $EQE_{LSC}$) borosilicate glass sheets and allowed to dry for 6h in a glove-box filled with nitrogen gas ($O_2$, $H_2O$ < 1ppm). After the composite films were completely dry, two components were encapsulated together by UV-curing epoxy (DELO) around the edges, where the two composite films faced each other within the encapsulation. The edge-mounted GaAs PVs (Alta Devices) were used as received. For *J-V* measurements, two PV strips were mounted on orthogonal edges (each edge was fully covered) using index matching gel (Thorlabs) to attach the PV strips on the waveguide edges and were connected in parallel. The remaining two edges were painted black to block the light and internal reflection of light. For $EQE_{LSC}$ measurements, one PV strip (composed of two GaAs PVs connected in parallel) was attached to one edge of the waveguide with the other three edges painted black. Correcting the raw data to account for 4 cell integration was done according to standardized protocols reported elsewhere.[28]

*Optical Characterization*:

Specular transmittance ($T(\lambda)$) of TLSC devices were measured using a double-beam Lambda 800 UV/VIS spectrometer in the transmission mode. No reference sample was placed on the reference beam side for the solid-film TLSC transmittance measurement. Reflectance ($R(\lambda)$) of the TLSCs was also measured using a Lambda 800 UV/VIS spectrometer with the 6◦ specular accessory installed on the sample beam side. The absorption spectra ($A(\lambda)$) were acquired by following the equation: $A(\lambda) = 1 - T(\lambda) - R(\lambda)$. The PL for NC, $CO_i8DFIC$ and BODIPY in polymer matrix were measured with a PTI QuantaMaster 40 spectrofluorometer with excitation at 400 nm, 650 nm and 680 nm, respectively. Photoluminescence quantum yields of NC, $CO_i8DFIC$ and BODIPY were measured using a Hamamatsu Quantaurus fluorometer.

*Module Photovoltaic Characterization*:

A Keithley 2420 SourceMeter was used to obtain *J-V* characteristics under simulated AM 1.5G solar illumination. A xenon arc lamp was used as the illumination source and the $EQE_{LSC}$ spectra of each TLSCs were used as the input to calculate their corresponding mismatch factors (*MF*): the *MF* values are 1.067, 1.051 and 1.052 for NC-only, NC+COi8DFIC and NC+BODIPY TLSCs. The light intensity was calibrated with an NREL-calibrated Si reference diode with a KG5 filter. The position-dependent $EQE_{LSC}$ measurements were performed using a QTH lamp with a calibrated Si detector, monochromator, chopper and lock-in amplifier. The measured $EQE_{LSC}(\lambda)$ at each distance (*d*) was corrected by multiplying the geometric factor $g = \pi/tan^{-1}(L/2d)$, which accounts for the different angle subtended by the edge-mounted PV at various excitation distance (*d*), where *L* is the LSC waveguide length. A series of $EQE_{LSC}(\lambda)$ spectra were acquired with the same TLSC device attached to the same GaAs PV, then the averaged spectrum was used to represent the whole device and integrated to confirm the $J_{SC}$ from the corresponding *J-V* characteristics of the same device. A matte black background was placed on the back of the TLSC device to eliminate illumination from the environment or reflection (double-pass) for both *J-V* and $EQE_{LSC}$ measurements. All the TLSC devices were tested with the same GaAs PV cells to eliminate any PV-to-PV variation in performance.

*Lifetime Test*:

A sulfur plasma lamp was used to constantly illuminate the TLSCs for photostability measurements. The illumination intensity of the lamp was calibrated to ~ 1 Sun with NREL-calibrated Si reference cell. Three key parameters including $A(\lambda)$, $EQE_{LSC}(\lambda)$ and $IQE_{LSC}(\lambda)$ were monitored to evaluate the photostability of the TLSC devices.

**References**

1. Traverse, C.J., Pandey, R., Barr, M.C., and Lunt, R.R. (2017). Emergence of highly transparent photovoltaics for distributed applications. Nat. Energy *2*, 849–860.

2. Currie, M.J., Mapel, J.K., Heidel, T.D., Goffri, S., and Baldo, M.A. (2008). High-Efficiency Organic Solar Concentrators for Photovoltaics. Science (80-. ). *321*, 226 LP – 228.

3. Yang, C., and Lunt, R.R. (2017). Limits of Visibly Transparent Luminescent Solar Concentrators. Adv. Opt. Mater. *5*, 1600851-n/a.

4. Debije, M.G., and Verbunt, P.P.C. (2012). Thirty years of luminescent solar concentrator research: Solar energy for the built environment. Adv. Energy Mater. *2*, 12–35.

5. Lunt, R.R. (2012). Theoretical limits for visibly transparent photovoltaics. Appl. Phys. Lett. *101*, 43902.

6. Chen, W., Zhang, J., Xu, G., Xue, R., Li, Y., Zhou, Y., Hou, J., and Li, Y. (2018). A Semitransparent Inorganic Perovskite Film for Overcoming Ultraviolet Light Instability of Organic Solar Cells and Achieving 14.03% Efficiency. Adv. Mater. *30*, 1800855.

7. Liu, D., Yang, C., and Lunt, R.R. (2018). Halide Perovskites for Selective Ultraviolet-Harvesting Transparent Photovoltaics. Joule *2*, 1827–1837.

8. Xiao, Z., Jia, X., Li, D., Wang, S., Geng, X., Liu, F., Chen, J., Yang, S., Russell, T.P., and Ding, L. (2017). 26 mA cm−2 Jsc from organic solar cells with a low-bandgap nonfullerene acceptor. Sci. Bull. *62*, 1494–1496.

9. Meng, L., Zhang, Y., Wan, X., Li, C., Zhang, X., Wang, Y., Ke, X., Xiao, Z., Ding, L., Xia, R., et al. (2018). Organic and solution-processed tandem solar cells with 17.3% efficiency. Science (80-. ). *361*, 1094–1098.

10. Li, Y., Ji, C., Qu, Y., Huang, X., Hou, S., Li, C.Z., Liao, L.S., Guo, L.J., and Forrest, S.R. (2019). Enhanced Light Utilization in Semitransparent Organic Photovoltaics Using an Optical Outcoupling Architecture. Adv. Mater. *31*, 1903173.

11. Zuo, L., Shi, X., Fu, W., and Jen, A.K.Y. (2019). Highly Efficient Semitransparent Solar Cells with Selective Absorption and Tandem Architecture. Adv. Mater. *31*, 1901683.

12. Xue, Q., Xia, R., Brabec, C.J., and Yip, H.L. (2018). Recent advances in semi-transparent polymer and perovskite solar cells for power generating window applications. Energy Environ. Sci. *11*, 1688–1709.

13. Wang, J., Zhang, J., Xiao, Y., Xiao, T., Zhu, R., Yan, C., Fu, Y., Lu, G., Lu, X., Marder, S.R., et al. (2018). Effect of Isomerization on High-Performance Nonfullerene Electron Acceptors. J. Am. Chem. Soc. *140*, 9140–9147.

14. Zuo, L., Shi, X., Jo, S.B., Liu, Y., Lin, F., and Jen, A.K.-Y. (2018). Tackling Energy Loss for High-Efficiency Organic Solar Cells with Integrated Multiple Strategies.

Adv. Mater. *30*, 1706816.

15. Yang, C., Moemeni, M., Bates, M., Sheng, W., Borhan, B., and Lunt, R.R. (2020). High-Performance Near-Infrared Harvesting Transparent Luminescent Solar Concentrators. Adv. Opt. Mater. *n/a*, 1901536.

16. Bergren, M.R., Makarov, N.S., Ramasamy, K., Jackson, A., Guglielmetti, R., and McDaniel, H. (2018). High-Performance CuInS2 Quantum Dot Laminated Glass Luminescent Solar Concentrators for Windows. ACS Energy Lett. *3*, 520–525.

17. Liu, C., and Li, B. (2015). Multiple dyes containing luminescent solar concentrators with enhanced absorption and efficiency. J. Opt. *17*, 25901.

18. Liu, G., Zhao, H., Diao, F., Ling, Z., and Wang, Y. (2018). Stable tandem luminescent solar concentrators based on CdSe/CdS quantum dots and carbon dots. J. Mater. Chem. C *6*, 10059–10066.

19. Mateen, F., Ali, M., Lee, S.Y., Jeong, S.H., Ko, M.J., and Hong, S.-K. (2019). Tandem structured luminescent solar concentrator based on inorganic carbon quantum dots and organic dyes. Sol. Energy *190*, 488–494.

20. Wu, K., Li, H., and Klimov, V.I. (2018). Tandem luminescent solar concentrators based on engineered quantum dots. Nat. Photonics *12*, 105–110.

21. Zhou, Y., Benetti, D., Tong, X., Jin, L., Wang, Z.M., Ma, D., Zhao, H., and Rosei, F. (2018). Colloidal carbon dots based highly stable luminescent solar concentrators. Nano Energy *44*, 378–387.

22. Ma, W., Li, W., Liu, R., Cao, M., Zhao, X., and Gong, X. (2019). Carbon dots and AIE molecules for highly efficient tandem luminescent solar concentrators. Chem. Commun. *55*, 7486–7489.

23. Desmet, L., Ras, A.J.M., de Boer, D.K.G., and Debije, M.G. (2012). Monocrystalline silicon photovoltaic luminescent solar concentrator with 4.2% power conversion efficiency. Opt. Lett. *37*, 3087–3089.

24. Zhao, H., Benetti, D., Tong, X., Zhang, H., Zhou, Y., Liu, G., Ma, D., Sun, S., Wang, Z.M., Wang, Y., et al. (2018). Efficient and stable tandem luminescent solar concentrators based on carbon dots and perovskite quantum dots. Nano Energy *50*, 756–765.

25. Banal, J.L., Zhang, B., Jones, D.J., Ghiggino, K.P., and Wong, W.W.H. (2017). Emissive Molecular Aggregates and Energy Migration in Luminescent Solar Concentrators. Acc. Chem. Res. *50*, 49–57.

26. Zhang, B., Banal, J.L., Jones, D.J., Tang, B.Z., Ghiggino, K.P., and Wong, W.W.H. (2018). Aggregation-induced emission-mediated spectral downconversion in luminescent solar concentrators. Mater. Chem. Front. *2*, 615–619.

27. Banal, J.L., Soleimaninejad, H., Jradi, F.M., Liu, M., White, J.M., Blakers, A.W., Cooper, M.W., Jones, D.J., Ghiggino, K.P., Marder, S.R., et al. (2016). Energy Migration in Organic Solar Concentrators with a Molecularly Insulated Perylene Diimide. J. Phys. Chem. C *120*, 12952–12958.

28. Yang, C., Liu, D., and Lunt, R.R. (2019). How to Accurately Report Transparent Luminescent Solar Concentrators. Joule *3*, 2871–2876.

29. Yang, C., Liu, D., Bates, M., Barr, M.C., and Lunt, R.R. (2019). How to Accurately Report Transparent Solar Cells. Joule *3*, 1803–1809.

30. Goetzberger, A., and Greube, W. (1977). Solar energy conversion with fluorescent collectors. Appl. Phys. *14*, 123–139.

31. Giebink, N.C., Wiederrecht, G.P., and Wasielewski, M.R. (2011). Resonance-shifting to circumvent reabsorption loss in luminescent solar concentrators. Nat. Photonics *5*, 694.

32. Walheim, S., Schäffer, E., Mlynek, J., and Steiner, U. (1999). Nanophase-Separated Polymer Films as High-Performance Antireflection Coatings. Science (80-. ). *283*, 520 LP – 522.

33. Sydlik, S.A., Chen, Z., and Swager, T.M. (2011). Triptycene Polyimides: Soluble Polymers with High Thermal Stability and Low Refractive Indices. Macromolecules *44*, 976–980.

34. Groh, W., and Zimmermann, A. (1991). What is the lowest refractive index of an organic polymer? Macromolecules *24*, 6660–6663.

35. Sarma, T., Panda, P.K., and Setsune, J. (2013). Bis-naphthobipyrrolylmethene derived BODIPY complex: an intense near-infrared fluorescent dye. Chem. Commun. *49*, 9806–9808.

36. Xi, J.-Q., Kim, J.K., Schubert, E.F., Ye, D., Lu, T.-M., Lin, S.-Y., and Juneja, J.S. (2006). Very low-refractive-index optical thin films consisting of an array of SiO2 nanorods. Opt. Lett. *31*, 601–603.

37. Xi, J.-Q., Schubert, M.F., Kim, J.K., Schubert, E.F., Chen, M., Lin, S.-Y., Liu, W., and Smart, J.A. (2007). Optical thin-film materials with low refractive index for broadband elimination of Fresnel reflection. Nat. Photonics *1*, 176.

38. Yan, X., Poxson, D.J., Cho, J., Welser, R.E., Sood, A.K., Kim, J.K., and Schubert, E.F. (2013). Enhanced Omnidirectional Photovoltaic Performance of Solar Cells Using Multiple-Discrete-Layer Tailored- and Low-Refractive Index Anti-Reflection Coatings. Adv. Funct. Mater. *23*, 583–590.

39. Schubert, E.F., Kim, J.K., and Xi, J.-Q. (2007). Low-refractive-index materials: A new class of optical thin-film materials. Phys. status solidi *244*, 3002–3008.

40. Yang, C., Liu, D., Renny, A., Kuttipillai, P.S., and Lunt, R.R. (2019). Integration of near-infrared harvesting transparent luminescent solar concentrators onto arbitrary surfaces. J. Lumin. *210*, 239–246.

41. Kirakci, K., Kubát, P., Dušek, M., Fejfarová, K., Šícha, V., Mosinger, J., and Lang, K. (2012). A Highly Luminescent Hexanuclear Molybdenum Cluster – A Promising Candidate toward Photoactive Materials. Eur. J. Inorg. Chem. *2012*, 3107–3111.

42. Saito, N., Cordier, S., Lemoine, P., Ohsawa, T., Wada, Y., Grasset, F., Cross, J.S., and Ohashi, N. (2017). Lattice and Valence Electronic Structures of Crystalline

Octahedral Molybdenum Halide Clusters-Based Compounds, Cs2[Mo6X14] (X = Cl, Br, I), Studied by Density Functional Theory Calculations. Inorg. Chem. *56*, 6234–6243.

43. Riehl, L., Ströbele, M., Enseling, D., Jüstel, T., and Meyer, H.-J. (2016). Molecular Oxygen Modulated Luminescence of an Octahedro-hexamolybdenum Iodide Cluster having Six Apical Thiocyanate Ligands. Zeitschrift für Anorg. und Allg. Chemie *642*, 403–408.

44. Kuttipillai, P.S., Zhao, Y., Traverse, C.J., Staples, R.J., Levine, B.G., and Lunt, R.R. (2016). Phosphorescent Nanocluster Light-Emitting Diodes. Adv. Mater. *28*, 320–326.

45. Kuttipillai, P.S., Yang, C., Chen, P., Wang, L., Bates, M., Lunt, S.Y., and Lunt, R.R. (2018). Enhanced Electroluminescence Efficiency in Metal Halide Nanocluster Based Light Emitting Diodes through Apical Halide Exchange. ACS Appl. Energy Mater. *1*, 3587–3592.

46. Zhao, Y., and Lunt, R.R. (2013). Transparent Luminescent Solar Concentrators for Large-Area Solar Windows Enabled by Massive Stokes-Shift Nanocluster Phosphors. Adv. Energy Mater. *3*, 1143–1148.

47. Yang, C., Zhang, J., Peng, W.T., Sheng, W., Liu, D., Kuttipillai, P.S., Young, M., Donahue, M.R., Levine, B.G., Borhan, B., et al. (2018). Impact of Stokes Shift on the Performance of Near-Infrared Harvesting Transparent Luminescent Solar Concentrators. Sci. Rep. *8*, 16359.

48. Zhao, H., Sun, R., Wang, Z., Fu, K., Hu, X., and Zhang, Y. (2019). Zero-Dimensional Perovskite Nanocrystals for Efficient Luminescent Solar Concentrators. Adv. Funct. Mater. *29*, 1902262.

49. Lee, K., Kim, N., Kim, K., Um, H.D., Jin, W., Choi, D., Park, J., Park, K.J., Lee, S., and Seo, K. (2020). Neutral-Colored Transparent Crystalline Silicon Photovoltaics. Joule *4*, 235–246.

50. Zhao, H., Zhou, Y., Benetti, D., Ma, D., and Rosei, F. (2017). Perovskite quantum dots integrated in large-area luminescent solar concentrators. Nano Energy *37*, 214–223.

**Acknowledgments**

**General**: The authors thank Dr. Dianyi Liu from School of Engineering, Westlake University for useful discussions. The authors also thank Scott Bankroff from MSU Department of Chemistry for preparing glass samples for this work.

**Funding:** National Science Foundation under grant CBET-1702591 and James Dyson Fellowship.

**Author contributions:**

C.Y., W.S., B.B. and R.R.L conceived and designed the project. C.Y. synthesized the nanoclusters, characterized the optical and photoluminescent properties of the luminophores, fabricated TLSC devices, and acquired device data. W.S. and M.M. synthesized the organic compounds ($CO_i$8DFIC and BODIPY) and purified the nanoclusters. M.B. assisted with the mass spectrometry of the nanoclusters and the photovoltaic measurements of the TLSC devices. C.K.H. assisted with the photostability measurements. C.Y. W.S. and R.R.L. wrote the manuscript.

**Competing interests:**

A provisional patent has been filed based on the work in this manuscript. R.R.L. is a founder, advisor, and a minority owner of Ubiquitous Energy, Inc., a company working to commercialize TPV technologies.

**Figures and Tables**

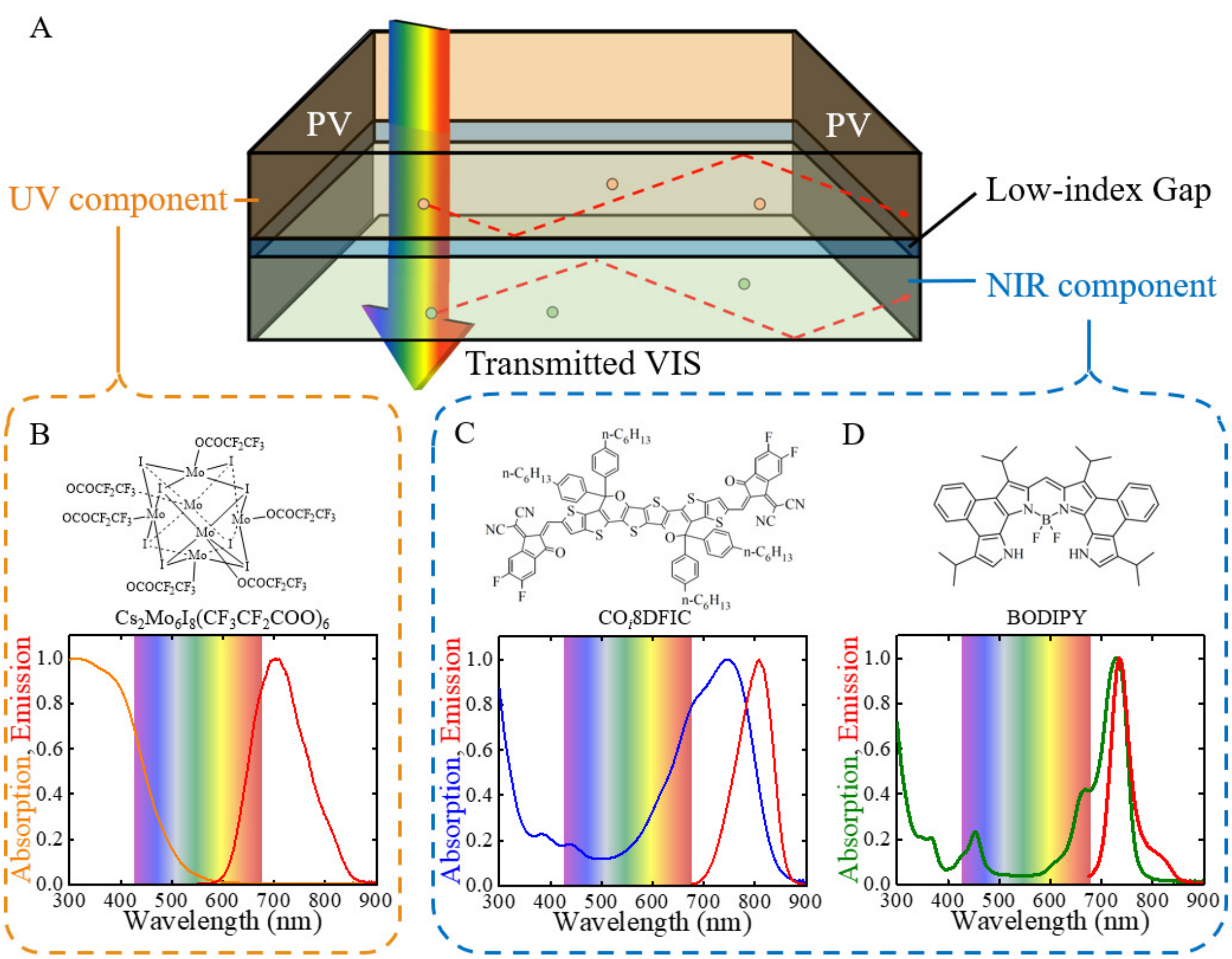

**Figure 1 Working Principle and Luminophores of the Dual-band TLSCs.** (**A**) Schematic showing the structure and working principle of the dual-band selective harvesting transparent luminescent solar concentrator (TLSC). The UV component and NIR component are separated by an air gap which enables total internal reflection within each waveguide and isolation of the emission from each luminophore. Molecular structure, normalized absorption and emission spectra of (**B**) $Cs_2Mo_6I_8(CF_3CF_2COO)_6$ nanocluster, (**C**) $CO_i8DFIC$ and (**D**) BODIPY in polymer matrix. Both the absorption and emission profiles of all the luminophores are designed to stay out of the VIS range, maximizing the visible transmission and aesthetic quality. Although a small portion of the NC PL falls into the red range, the majority of the escaped PL from the top UV component can be reabsorbed by the bottom NIR component, enhancing the light harvesting and minimizing any red glow observed from the transmitted side.

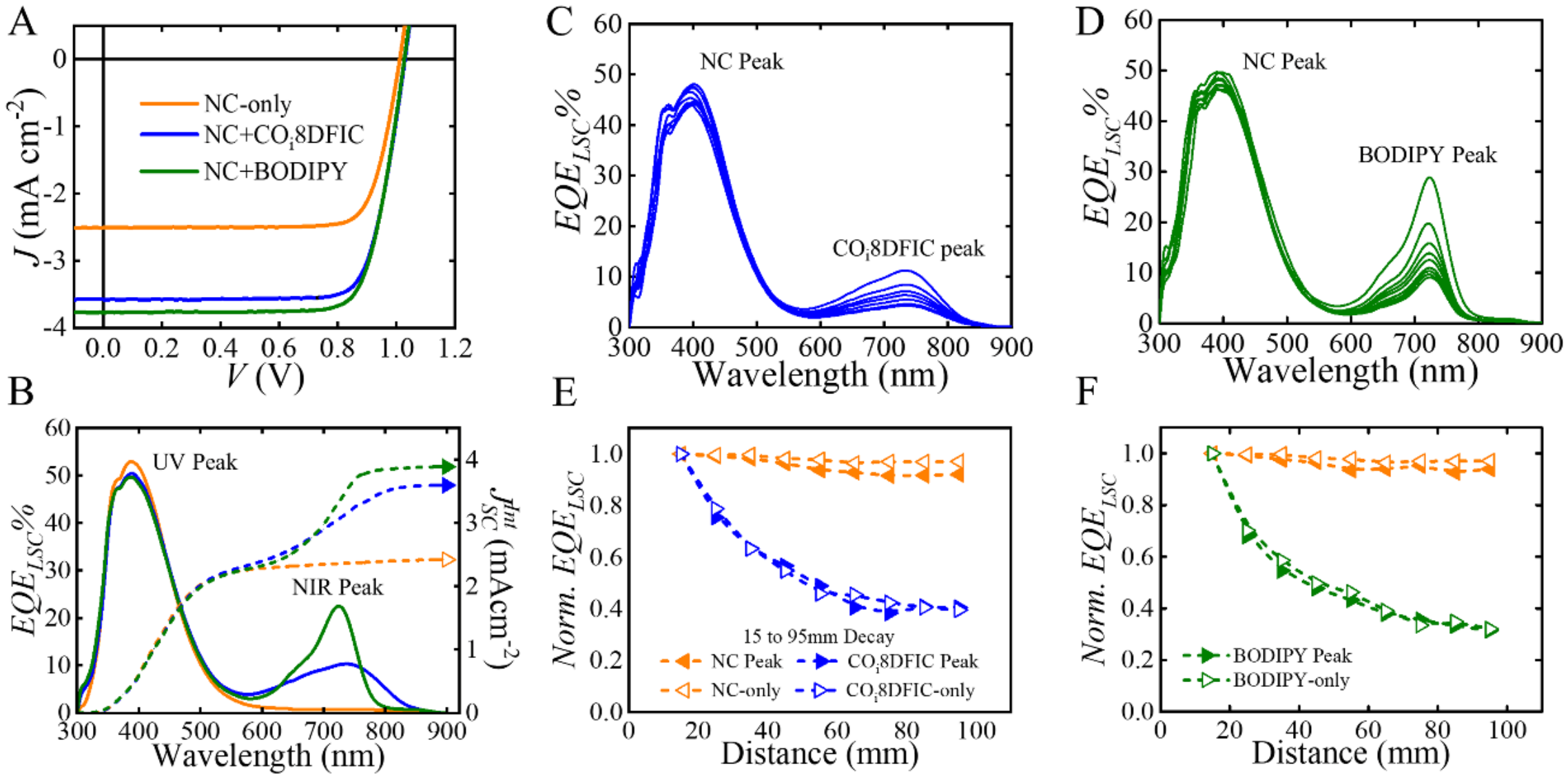

**Figure 2 Photovoltaic Performance of the Dual-band TLSCs.** (**A**) Current density versus voltage (*J-V*) characteristics of NC-only, NC+$CO_i$8DFIC and NC+BODIPY TLSCs. All scans were measured under AM 1.5G illumination and all TLSCs were edge-mounted with the same GaAs PV cells. (**B**) Average $EQE_{LSC}(\lambda)$ spectra of NC-only, NC+$CO_i$8DFIC and NC+BODIPY TLSCs. The corresponding integrated short-circuit current density ($J_{SC}^{Int}$) matches well with the $J_{SC}$ extracted from *J-V* characteristics shown in (A). The series of absolute position-dependent $EQE_{LSC}$ spectra of (**C**) NC+$CO_i$8DFIC and (**D**) NC+BODIPY TLSCs, where $d$ increases from 15 mm to 95 mm with 10 mm interval. The position-dependent $EQE_{LSC}$ peak values of NC+$CO_i$8DFIC and NC+BODIPY TLSCs are extracted, normalized and plotted in (**E**) and (**F**), respectively. A NC-only, a $CO_i$8DFIC-only and a BODIPY-only TLSC are included as references. the UV and NIR peak decay behaviors of the dual-band TLSCs closely resemble those of the NC-only, $CO_i$8DFIC-only and BODIPY-only reference TLSCs, respectively, confirming the effective isolation of the two components by the air gap.

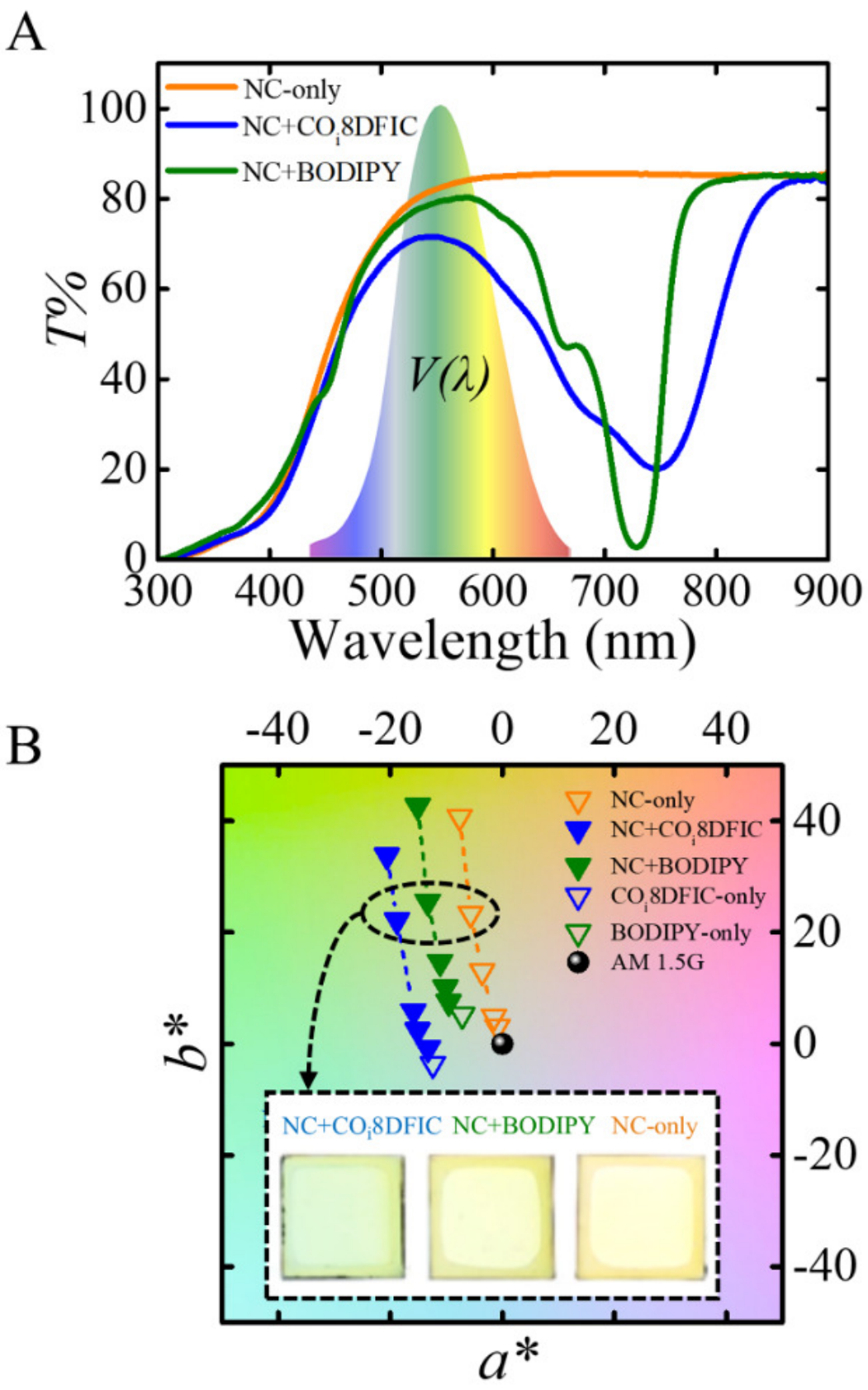

**Figure 3 Aesthetic Quality of the Dual-band TLSCs.** (**A**) The transmittance spectra (*T(λ)*) of the NC-only, NC+CO$_i$8DFIC and NC+BODIPY TLSCs along with the normalized photopic response of the human eye (*V(λ)*) for comparison. (**B**) The (*a**, *b**) coordinates of NC-only group, NC+CO$_i$8DFIC group and NC+BODIPY group TLSCs in CIELAB color space. Within each group the NC concentration (1, 2, 5, 10 and 20 mgmL$^{-1}$) is the only variable. The (*a**, *b**) of CO$_i$8DFIC-only and BODIPY-only TLSCs are included as references. Inset: photographs of NC-only, NC+CO$_i$8DFIC and NC+BODIPY TLSCs with NC concentration at 10 mgmL$^{-1}$.

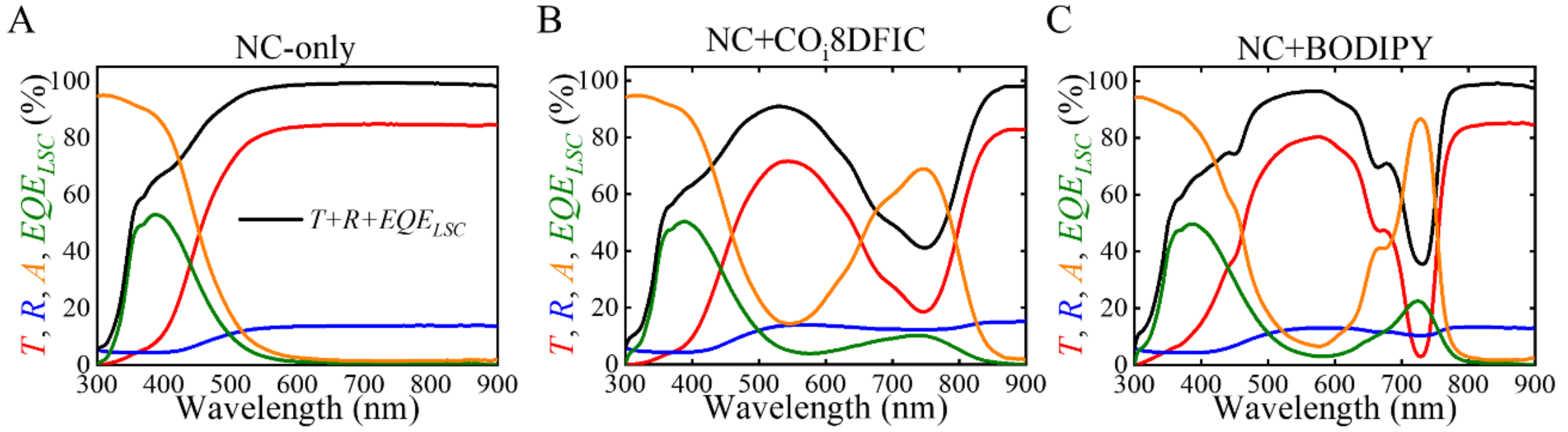

**Figure 4 Photon balance check.** (**A**) NC-only, (**B**) NC+CO$_i$8DFIC and (**C**) NC+BODIPY TLSCs.

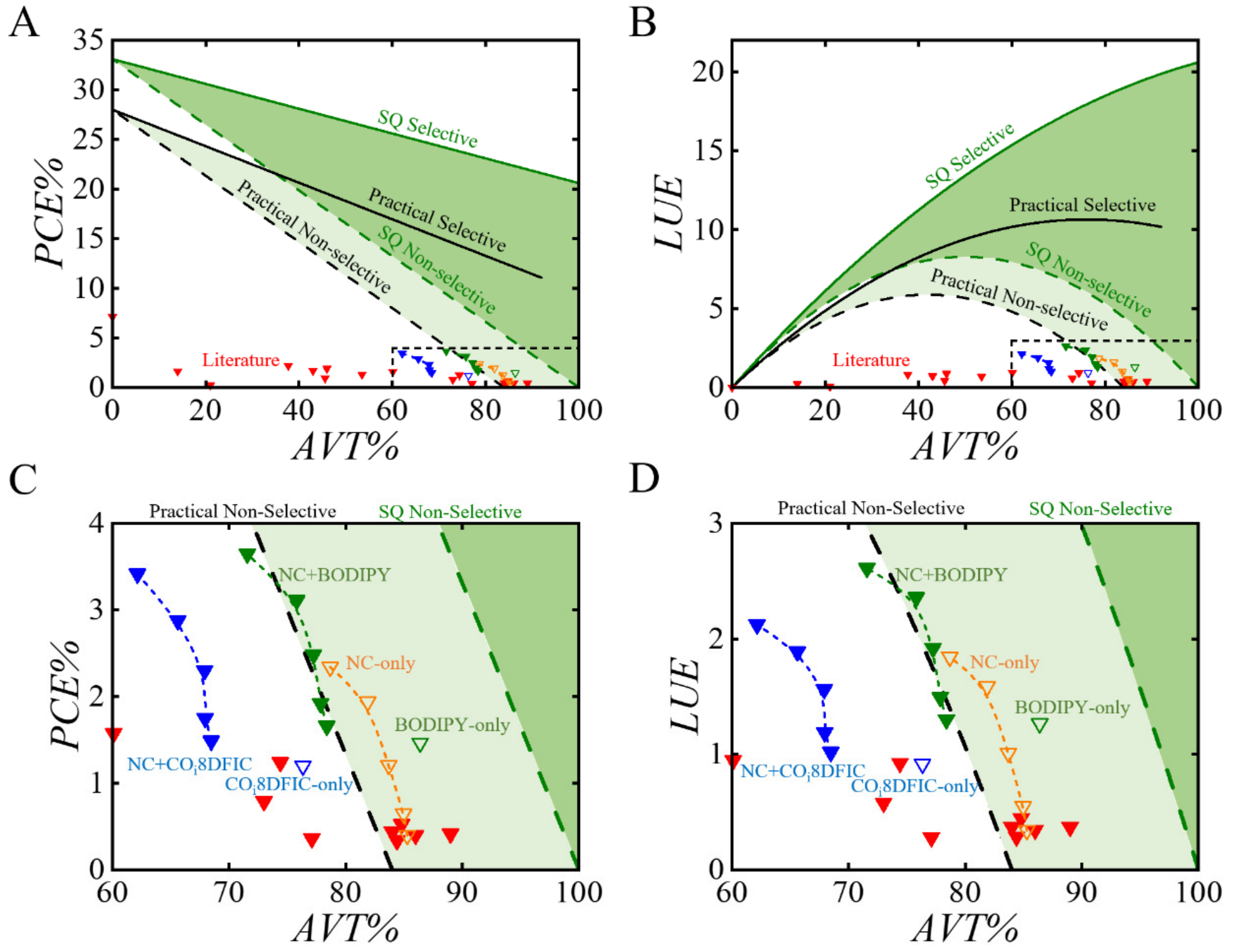

**Figure 5 Comprehensive Analysis of Photovoltaic Performance and Aesthetic Quality.** (**A**) Power conversion efficiency (*PCE*) versus average visible transmittance (*AVT*) and (**B**) Light utilization efficiency ($LUE = PCE \times AVT$) versus *AVT* for NC-only group, NC+$CO_i$8DFIC group, NC+BODIPY group, $CO_i$8DFIC-only and BODIPY-only TLSCs in full-scale. (**C**) Zoomed-in *PCE* vs. *AVT* plot and (D) zoomed-in *LUE* vs. *AVT* plot for all the TLSCs. Note: The olive dash line is the Shockley-Queisser (SQ) *PCE* (or *LUE*) limit for non-wavelength-selective TPV with partial visible transmittance; the black dashed line is the practical *PCE* (or *LUE*) limit for non-wavelength-selective TPV with partial visible transmittance. The dark shaded green region indicates the target *PCE* and *AVT* (or *LUE* and *AVT*) combination only achievable with the wavelength-selective approach (theoretical). The light shaded green region indicates the target *PCE* and *AVT* (or *LUE* and *AVT*) combination only achievable with the wavelength-selective approach (practical limits). Literature reports (red solid triangles, also tabulated in Supplemental Information Note 6 **Table S2**) are included in both plots for comparison. The dashed boxes at the bottom right corners of (**A**) and (**B**) are the zoomed-in scale for (**C**) and (**D**).

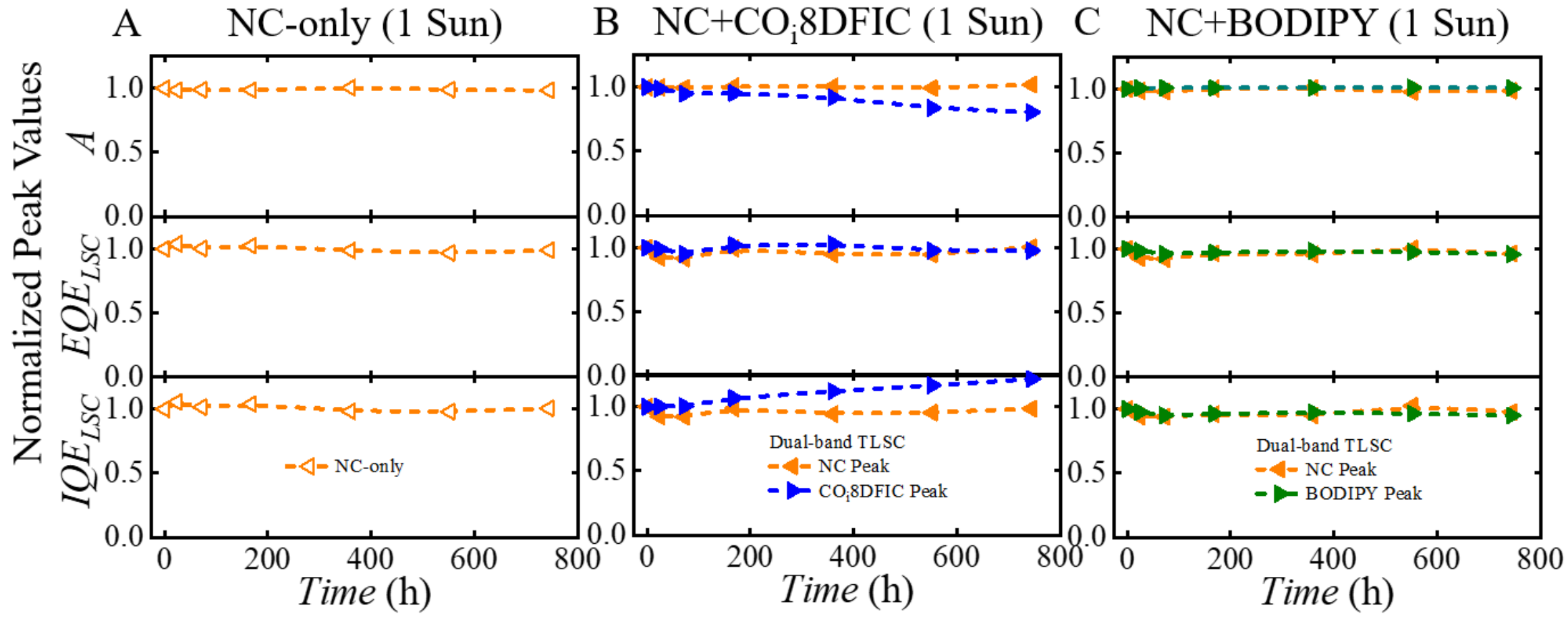

**Figure 6 Photostability study of dual-band TLSCs.** Normalized peak values of absorption spectra (*A(λ)*), $EQE_{LSC}(\lambda)$ and $IQE_{LSC}(\lambda)$ for (**A**) NC-only, (**B**) NC+$CO_i$8DFIC and (**C**) NC+BODIPY TLSCs as a function of time under constant illumination.

Supplemental Information for

# Ultraviolet and Near-infrared Dual-Band Selective-Harvesting Transparent Luminescent Solar Concentrators

Chenchen Yang,[1,†] Wei Sheng,[2,†] Mehdi Moemeni,[2] Matthew Bates,[1] Christopher K. Herrera,[1] Babak Borhan,[2] Richard R. Lunt[1,3,*]

[1] Department of Chemical Engineering and Materials Science, Michigan State University, East Lansing, MI, 48824 USA

[2] Department of Chemistry, Michigan State University, East Lansing, MI, 48824 USA

[3] Department of Physics and Astronomy, Michigan State University, East Lansing, MI, 48824 USA

* Correspondence: rlunt@msu.edu

† The authors C. Y. and W. S. contributed equally to this work.

**Note 1 X-ray Diffraction Pattern of $Cs_2Mo_6I_{14}$ Nanocluster Powder**

The observed X-ray diffraction (XRD) pattern of $Cs_2Mo_6I_{14}$ nanocluster (NC) shown in **Figure S1** matches well with previous literature report,[1] which confirms the formation of $Cs_2Mo_6I_{14}$ nanocluster from the synthesis of $MoI_2$ and CsI.

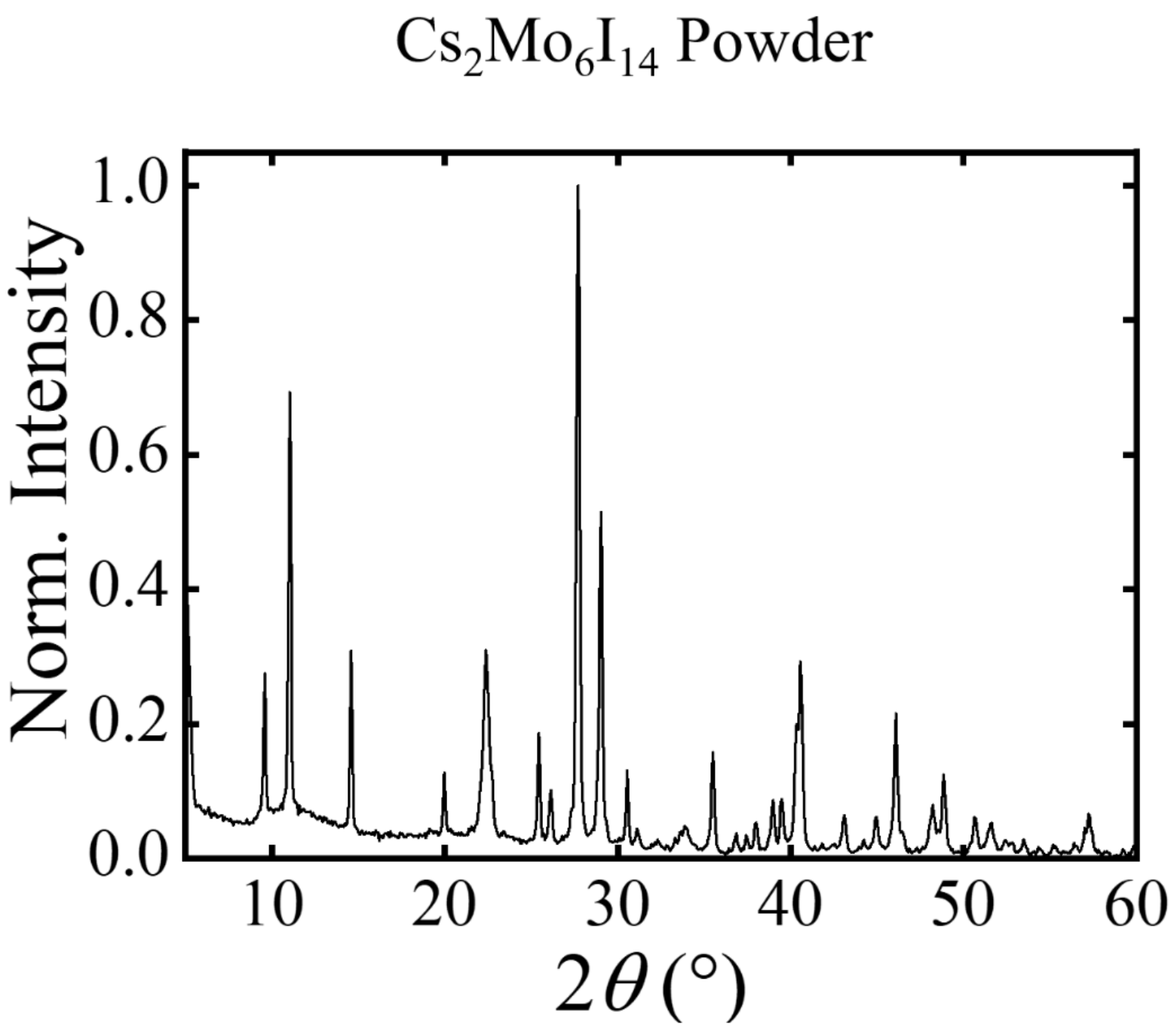

**Figure S1.** X-ray Diffraction pattern of $Cs_2$Mo6$I_{14}$ nanocluster powder.

## Note 2 Mass Spectrometry of Various Nanoclusters

High resolution mass spectrometry scans as a function of *m*/*z*[2,3] were measured and plotted in **Figure S2**. Mo has rich isotope distribution ($^{92}Mo$, $^{94}Mo$, $^{95}Mo$, $^{96}Mo$, $^{97}Mo$ and $^{98}Mo$ are the six main and stable isotopes of Mo), and there are six Mo sites in each NC. Various combination of these isotopes therefore results in the distribution in the corresponding mass spectrometry plot. The experimentally measured mass spectrometry patterns of the $Cs_2Mo_6I_{14}$, $Cs_2Mo_6I_8(CF_3COO)_6$, $Cs_2Mo_6I_8(CF_3CF_2COO)_6$ and $Cs_2Mo_6I_8(CF_3CF_2CF_2COO)_6$ NCs match well with theoretical peaks, confirming the successful synthesis of $CsMo_6I1_4$ NC and subsequent substitution of the apical halide positions with various ligands (including $CF_3COO^-$, $CF_3CF_2COO^-$, and $CF_3CF_2CF_2COO^-$).

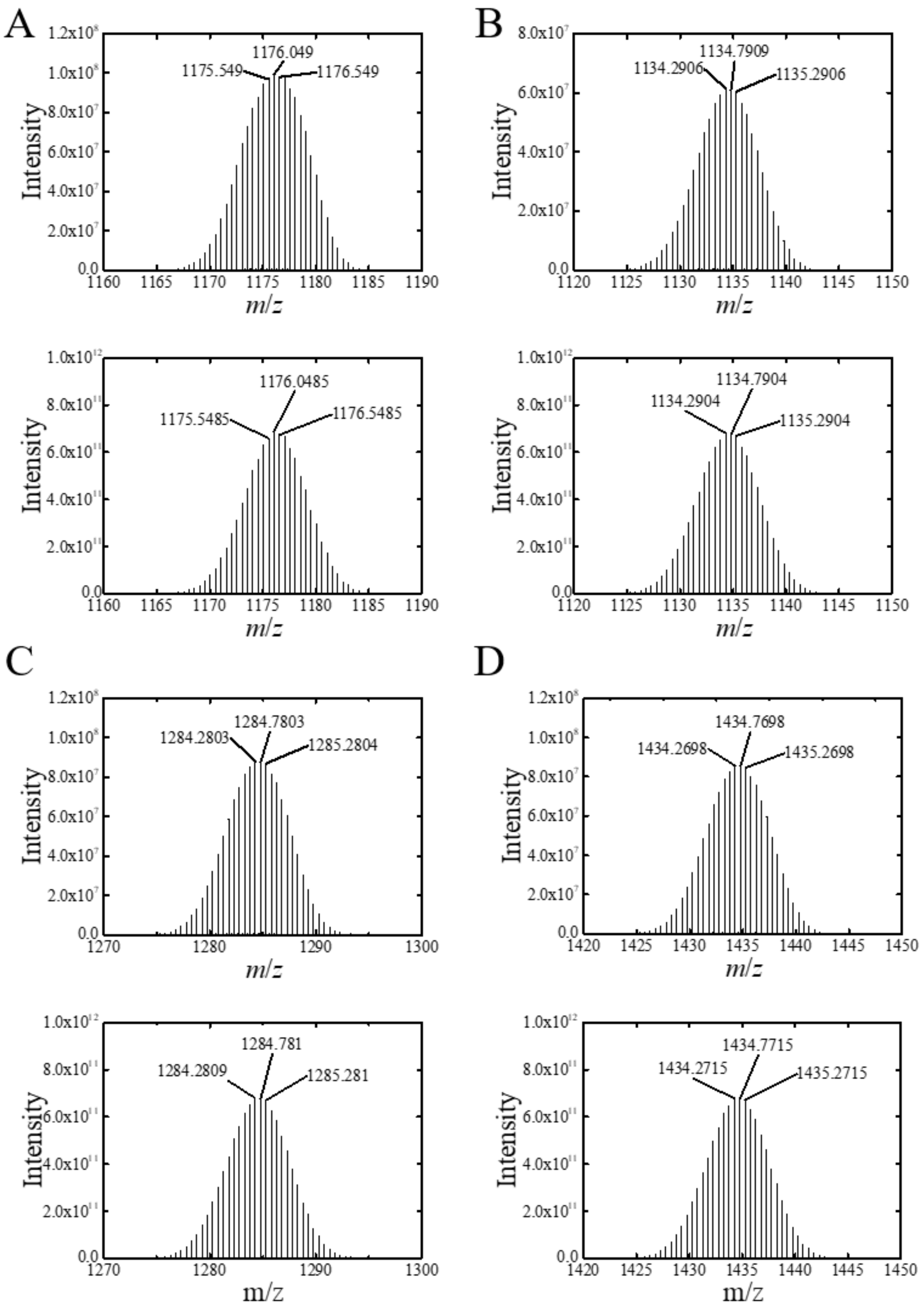

**Figure S2.** Mass spectrometry patterns of (A) $Cs_2Mo_6I_{14}$, (B) $Cs_2Mo_6I_8(CF_3COO)_6$, (C) $Cs_2Mo_6I_8(CF_3CF_2COO)_6$ and (D) $Cs_2Mo_6I_8(CF_3CF_2CF_2COO)_6$ NCs with the experimental measured data (top) compared with the theoretical isotopic distribution (bottom) in each plot.

## Note 3 $CO_i$8DFIC-only and BODIPY-only TLSCs

The NIR component only ($CO_i$8DFIC-only and BODIPY-only) TLSCs (**Figure S3**A) were characterized for reference in this work.

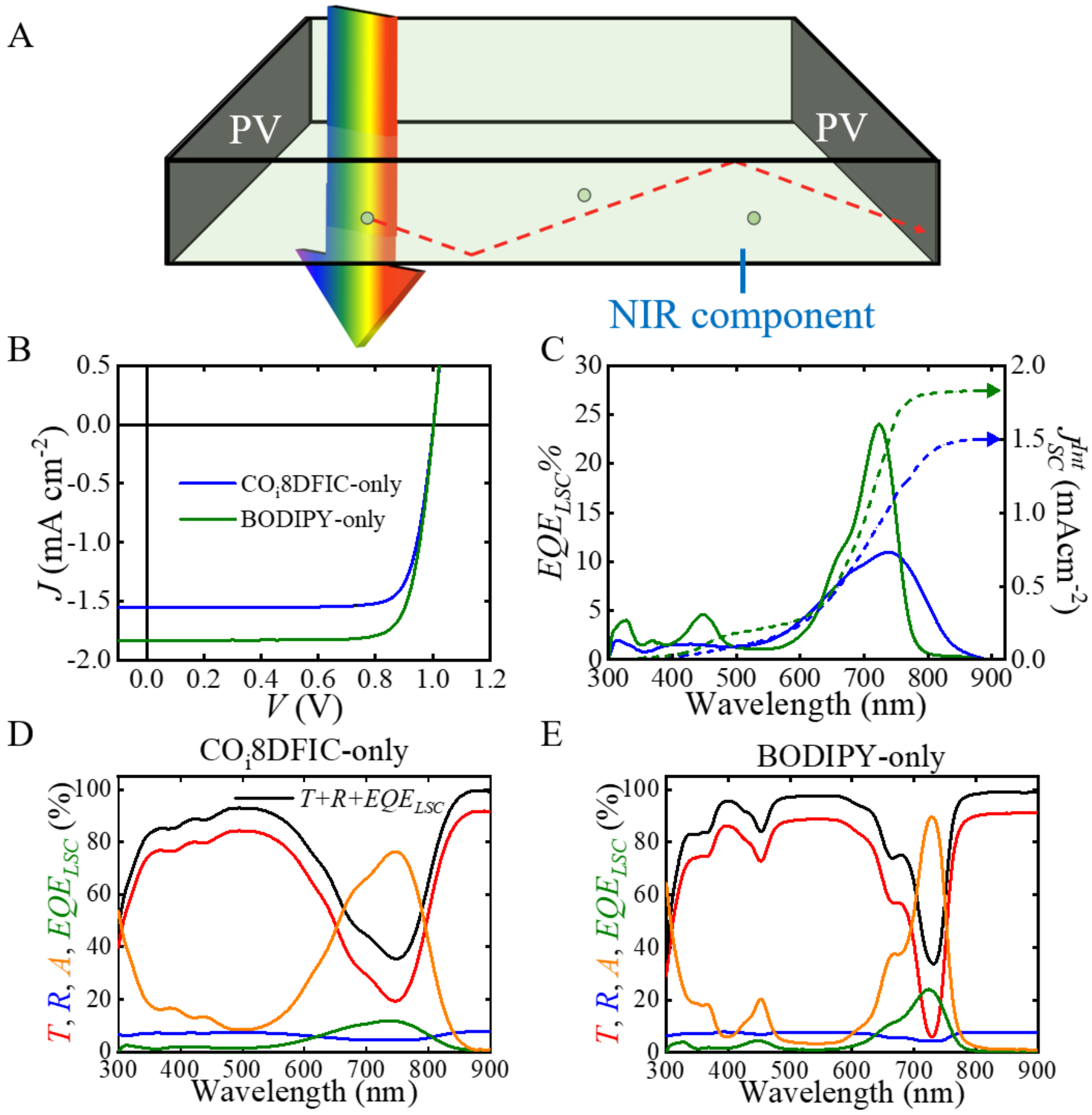

**Figure S3.** (A) Schematic of NIR component only TLSC. (B) Current density versus voltage (*J*-*V*) characteristics of $CO_i$8DFIC-only and BODIPY-only TLSCs under AM 1.5G illumination. (C) Average external quantum efficiency ($EQE_{LSC}(\lambda)$) spectra of $CO_i$8DFIC-only and BODIPY-only TLSCs. The corresponding integrated short-circuit current density ($J_{SC}^{Int}$) match well with the $J_{SC}$ extracted from *J*-*V* characteristics shown in (B). Photon balance check for (D) $CO_i$8DFIC-only and (E) BODIPY-only TLSCs.

As shown in Figure S3B the $CO_i8DFIC$-only TLSC shows short-circuit current density ($J_{SC}$) of 1.55±0.04 $mAcm^{-2}$, open-circuit voltage ($V_{OC}$) of 1.00±0.01 V and fill factor ($FF$) of 81±1%, resulting in a power conversion efficiency ($PCE$) of 1.26±0.03%. With significantly higher quantum yield ($QY$) of BODIPY compared to $CO_i8DFIC$, the BODIPY-only TLSC shows improved $J_{SC}$ of 1.84±0.03 $mAcm^{-2}$ with similar $V_{OC}$ and $FF$, leading to a corresponding $PCE$ of 1.48±0.03%. The $J_{SC}$ values extracted from current density versus voltage ($J$-$V$) characteristics are confirmed by the integrated $J_{SC}$ ( $J_{SC}^{Int}$ ) from the external quantum efficiency of LSC ($EQE_{LSC}(\lambda)$) as shown in Figure S3C. The peaks of $EQE_{LSC}$ match with the peaks of the absorption spectra of the corresponding organic luminophores, and the $J_{SC}^{Int}$ values are 1.50 $mAcm^{-2}$ and 1.83 $mAcm^{-2}$ for $CO_i8DFIC$-only and BODIPY-only TLSCs, respectively, which are in good agreement of the $J_{SC}$ values from the $J$-$V$ curves. The photon balance for $CO_i8DFIC$-only and BODIPY-only TLSCs is consistent ($EQE_{LSC}(\lambda)+T(\lambda)+R(\lambda) \leq 1$) as shown in Figure S3D and E, respectively.

**Note 4 CO$_i$8DFIC+NC and BODIPY+NC TLSCs**

Since the absorption profiles of the UV and NIR components are spectrally separated from each other, switching the sequence of the incident light passing through (NIR component as the top waveguide and UV component as the bottom waveguide as shown in **Figure S4**A) can still maintain good photovoltaic performance, which maintains the same aesthetic quality of the TLSC observed from the transmitted side.

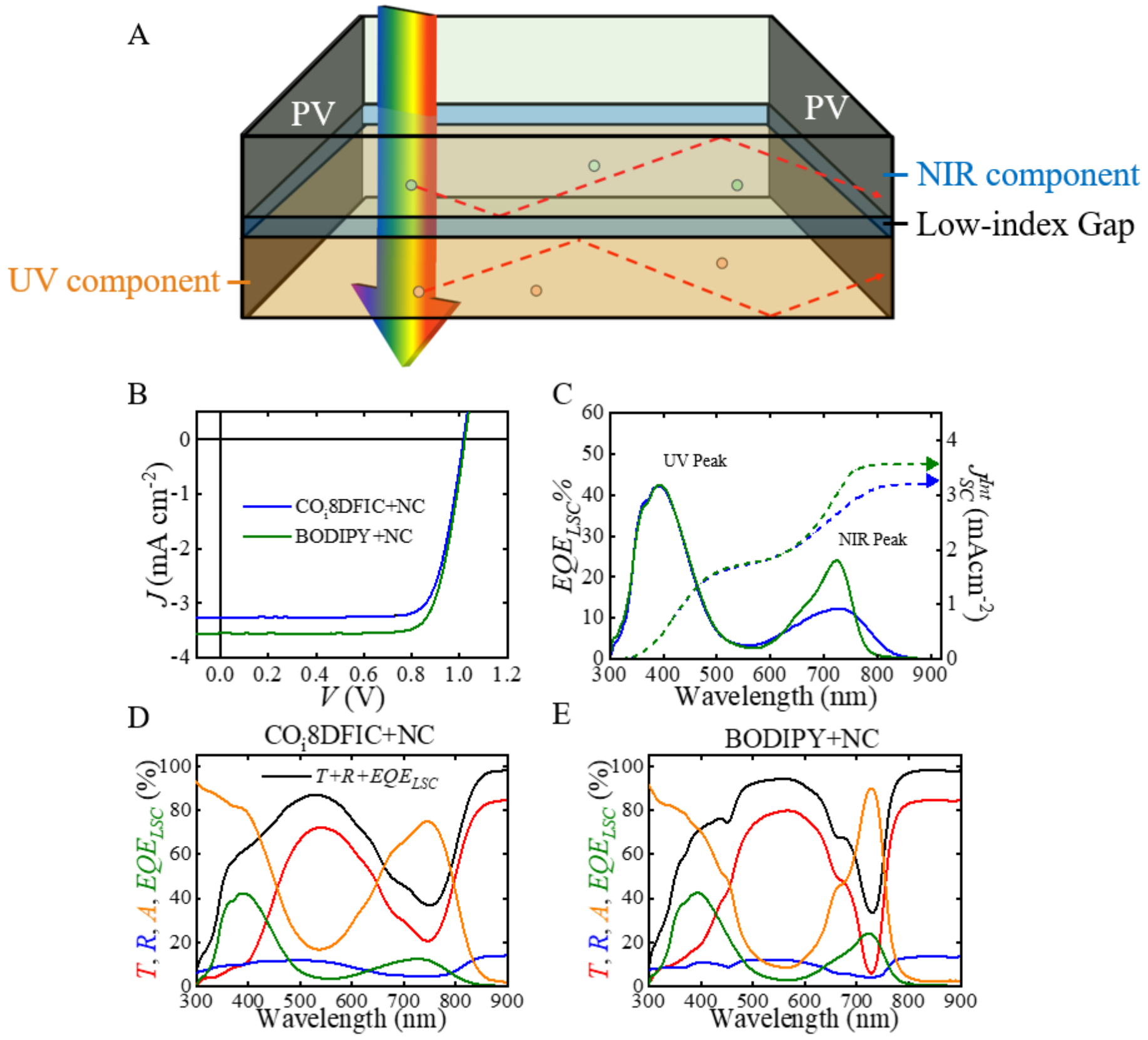

**Figure S4.** (A) Schematic showing the dual-band TLSCs with NIR component as the top waveguide and UV component as the bottom waveguide. (B) *J-V* characteristics of CO$_i$8DFIC+NC and BODIPY+NC TLSCs under AM 1.5G illumination. (C) Average $EQE_{LSC}(\lambda)$ spectra of CO$_i$8DFIC+NC and BODIPY+NC TLSCs. The corresponding integrated short-circuit current density ( $J_{SC}^{Int}$ ) match well with the $J_{SC}$ extracted from *J-V* characteristics shown in (B). Photon balance check for (D) CO$_i$8DFIC+NC and (E) BODIPY+NC TLSCs.

We note that in $CO_i8DFIC$+NC and BODIPY+NC TLSCs: $CO_i8DFIC$ or BODIPY is used as the top waveguide luminophore and NC is used as the bottom waveguide luminophore, and all the luminophore concentrations are kept the same as the NC+$CO_i8DFIC$ and NC+BODIPY TLSCs shown in **Figure 2**. In Figure S4B the $CO_i8DFIC$+NC TLSC shows $J_{SC}$ of 3.26±0.03 $mAcm^{-2}$, $V_{OC}$ of 1.01±0.01 V and *FF* of 79±1%, resulting in a *PCE* of 2.59±0.01%. With slightly higher $J_{SC}$ of 3.55±0.06 $mAcm^{-2}$ and similar $V_{OC}$ and *FF* values, the BODIPY+NC TLSC shows a *PCE* of 2.84±0.05%. Figure S4C shows the $EQE_{LSC}(\lambda)$ spectra of these two TLSCs, compared to Figure 2B the NC peaks decrease by ~10% due to more reflection loss of the UV photons, and the $CO_i8DFIC$ and BODIPY peaks increase by ~10% resulting from less reflection loss of the NIR photons. Therefore, the contribution to the overall $J_{SC}^{Int}$ from UV and NIR ranges becomes more balanced. The $J_{SC}^{Int}$ values of the $CO_i8DFIC$+NC and BODIPY+NC TLSCs are 3.26 $mAcm^{-2}$ and 3.57 $mAcm^{-2}$, which are in great agreement of the $J_{SC}$ values extracted from the corresponding *J-V* characteristics. The photon balance for $CO_i8DFIC$+NC and BODIPY+NC TLSCs is consistent as shown in Figure S4D and E, respectively.

Although the *PCE*s of $CO_i8DFIC$+NC and BODIPY+NC TLSCs are slightly lower than those of the NC+$CO_i8DFIC$ and NC+BODIPY TLSCs shown in Figure 2A, moving forward, with improved spectral coverage, *QY* and distinct separation of the absorption and emission spectra of the NIR selective-harvesting luminophores, the advantage of placing NIR component as the top waveguide could become more impactful and lead to superior *PCE* with the same aesthetic quality. However, as we note below it is also important to consider the impact of panel arrangement on lifetime, as putting the NC panel first can eliminate the UV from reach the NIR panel and in some cases help to extend the lifetime.

**Note 5 Photostability of the rest TLSCs**

Normalized peak values of absorption (*A(λ)*), $EQE_{LSC}(\lambda)$ and internal quantum efficiency ($IQE_{LSC}(\lambda) = EQE_{LSC}(\lambda)/A(\lambda)$) spectra for NC-only ($Cs_2Mo_6I_8(CF_3COO)_6$ and $Cs_2Mo_6I_8(CF_3CF_2CF_2COO)_6$), $CO_i8DFIC$+NC and BODIPY+NC are extracted and plotted as a function of time under constant illumination in **Figure S5**. The lifetime of an LSC is directly a function of the absorption efficiency (bleaching) of the luminophore, quantum yield of the luminophore, and lifetime of the edge-mounted PV. Since we are utilizing edge-mounted PVs with lifetimes of greater than 20 years we track the absorption profile and quantum efficiency of each luminophore combination.

All three parameters of all the NC-only (with various ligands) and the UV components of both dual-band TLSCs remain nearly constant after 700 hours of constant illumination. In BODIPY+NC TLSC the BODIPY peak also does not show any significant degradation. However, in $CO_i8DFIC$+NC TLSC with the NIR component as the top waveguide, the $CO_i8DFIC$ is not protected by the NC from the UV light, a more pronounced *A(λ)* decay of the $CO_i8DFIC$ is observed compared to that of the NC+$CO_i8DFIC$ TLSC. Compared to *A(λ)*, the $EQE_{LSC}(\lambda)$ of the $CO_i8DFIC$ peak shows a less pronounced decay trend due to less reabsorption loss, therefore, the corresponding $IQE_{LSC}(\lambda)$ even slightly increases at this time scale.

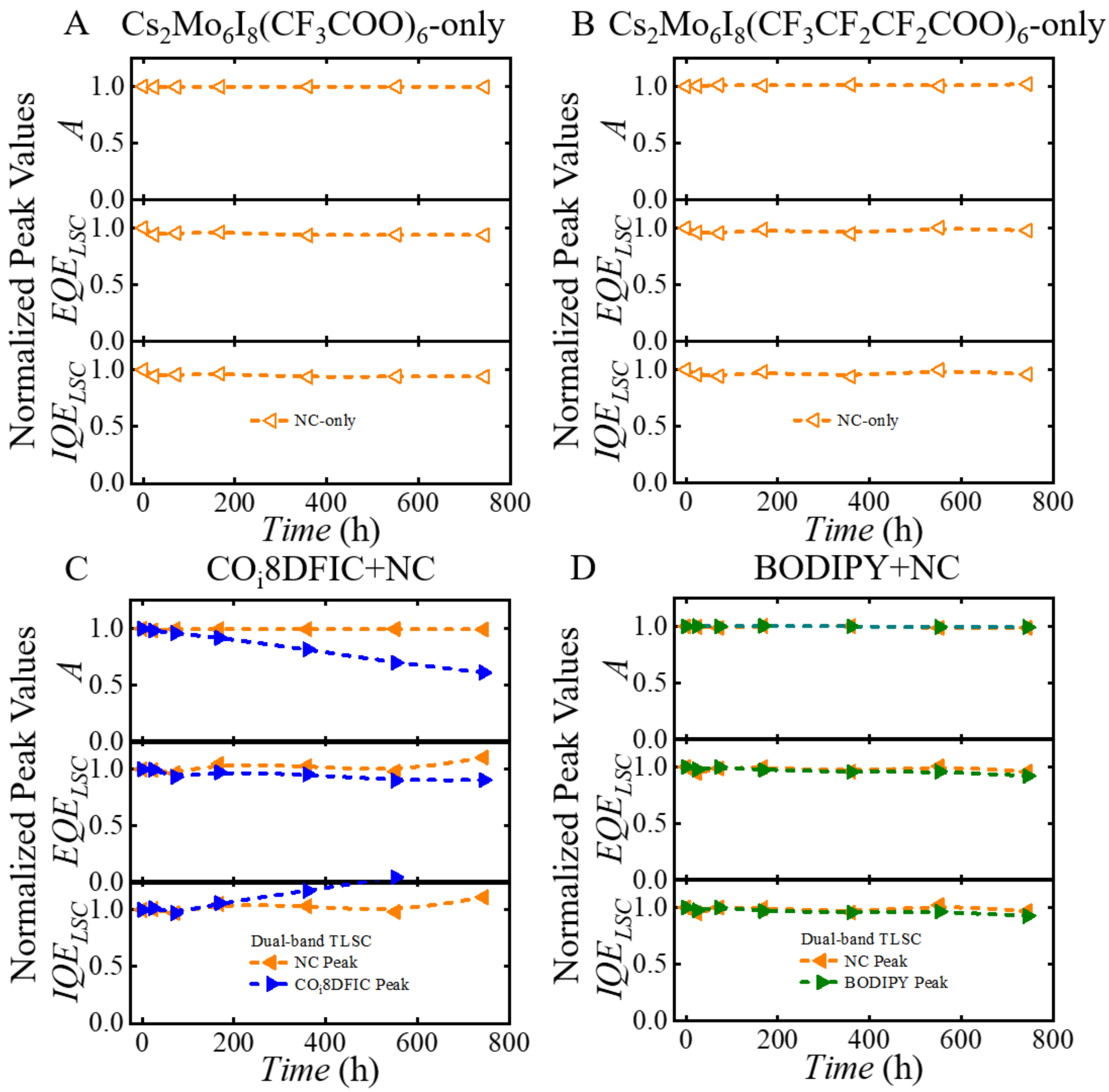

**Figure S5** Photostability study of dual-band TLSCs: normalized peak values of absorption spectra ($A(\lambda)$), $EQE_{LSC}(\lambda)$ and internal quantum yield ($IQE_{LSC}(\lambda)$) for (A) NC ($Cs_2Mo_6I_8(CF_3COO)_6$)-only, (B) NC ($Cs_2Mo_6I_8(CF_3CF_2CF_2COO)_6$)-only and (C) $CO_i8DFIC$+NC ($Cs_2Mo_6I_8(CF_3CF_2COO)_6$) and BODIPY+NC ($Cs_2Mo_6I_8(CF_3CF_2COO)_6$) TLSCs as a function of time under constant 1 Sun illumination.

## Note 6 Summary of All the TLSC Parameters and Literature Reports

**Table S1.** Photovoltaic and aesthetic quality parameters of the TLSCs.

| Luminophore(s) | *G* | $J_{SC}$ (mAcm$^{-2}$) | *Int.* $J_{SC}$ (mAcm$^{-2}$) | $V_{OC}$ (V) | *FF* % | *PCE* % | *AVT* % | *LUE* | *CRI* | (*a**, *b**) |
|---|---|---|---|---|---|---|---|---|---|---|
| NC (1)[a]-only | 4 | -[c] | 0.50 | 1.00[d] | 80[e] | 0.40 | 85.3 | 0.34 | 96.5 | (-0.8, 3.0) |
| NC (2)[a] -only | 4 | -[c] | 0.81 | 1.00[d] | 80[e] | 0.64 | 85.0 | 0.55 | 98.3 | (-1.5. 4.9) |
| NC (5)[a] -only | 4 | -[c] | 1.51 | 1.00[d] | 80[e] | 1.21 | 83.7 | 1.01 | 95.3 | (-3.6, 13.0) |
| NC (10)[a] -only | 4 | 2.0±0.1 | 2.42 | 1.01±0.01 | 80±1 | 1.94 | 81.9 | 1.59 | 91.3 | (-5.6, 23.3) |
| | 8[b] | -[c] | 2.10 | 1.00[d] | 80[e] | 1.68 | | 1.38 | | |
| NC (20)[a] -only | 4 | -[c] | 2.93 | 1.00[d] | 80[e] | 2.34 | 78.7 | 1.84 | 84.0 | (-7.6, 40.7) |
| NC (1)[a]+CO$_i$8DFIC | 2 | -[c] | 1.86 | 1.00[d] | 80[e] | 1.49 | 68.5 | 1.02 | 81.5 | (-13.3, -0.8) |
| NC (2)[a]+CO$_i$8DFIC | 2 | -[c] | 2.18 | 1.00[d] | 80[e] | 1.75 | 68.0 | 1.19 | 80.8 | (-15.1, 2.4) |
| NC (5)[a]+CO$_i$8DFIC | 2 | -[c] | 2.88 | 1.00[d] | 80[e] | 2.30 | 67.9 | 1.56 | 81.8 | (-15.8, 5.8) |
| NC (10)[a]+CO$_i$8DFIC | 2 | 3.6±0.2 | 3.60 | 1.02±0.01 | 79±1 | 2.9±0.1 | 65.6 | 1.89 | 82.9 | (-18.8, 22.2) |
| | 4[b] | -[c] | 3.00 | 1.00[d] | 80[e] | 2.40 | | 1.58 | | |
| CO$_i$8DFIC+NC (10)[a] | 2 | 3.26±0.03 | 3.26 | 1.01±0.01 | 79±1 | 2.59±0.01 | 66.4 | 1.72 | 84.2 | (-18.0, 22.9) |
| NC (20)[a]+CO$_i$8DFIC | 2 | -[c] | 4.27 | 1.00[d] | 80[e] | 3.42 | 62.1 | 2.12 | 80.4 | (-20.6, 33.7) |
| NC (1)[a] +BODIPY | 2 | -[c] | 2.07 | 1.00[d] | 80[e] | 1.66 | 78.4 | 1.30 | 89.8 | (-9.5, 7.5) |
| NC (2)[a] +BODIPY | 2 | -[c] | 2.40 | 1.00[d] | 80[e] | 1.92 | 77.9 | 1.49 | 90.1 | (-10.1. 10.1) |
| NC (5)[a] +BODIPY | 2 | -[c] | 3.10 | 1.00[d] | 80[e] | 2.48 | 77.2 | 1.92 | 90.2 | (-11.1, 14.7) |
| NC (10)[a] +BODIPY | 2 | 3.8±0.1 | 3.89 | 1.02±0.01 | 78±1 | 3.01±0.07 | 75.8 | 2.36 | 88.3 | (-13.3, 25.5) |
| | 4[b] | -[c] | 3.32 | 1.00[d] | 80[e] | 2.66 | | 2.02 | | |
| BODIPY+NC (10)[a] | 2 | 3.55±0.06 | 3.57 | 1.02±0.01 | 79±1 | 2.84±0.05 | 73.4 | 2.08 | 86.1 | (-15.4, 28.4) |
| NC (20)[a] +BODIPY | 2 | -[c] | 4.56 | 1.00[d] | 80[e] | 3.65 | 71.6 | 2.61 | 82.9 | (-15.1, 42.7) |
| CO$_i$8DFIC-only | 4 | 1.55±0.04 | 1.50 | 1.00±0.01 | 81±1 | 1.26±0.03 | 76.3 | 0.92 | 81.6 | (-12.4, -3.7) |
| BODIPY-only | 4 | 1.84±0.03 | 1.83 | 1.00±0.01 | 81±1 | 1.48±0.03 | 86.4 | 1.26 | 92.2 | (-7.2, 5.2) |

[a] Inside each () is the concentration of NC (in mgmL$^{-1}$) in the precursor solution.
[b] 10.16×10.16 cm$^2$ TLSCs for position-dependent $EQE_{LSC}$ roll-off behavior study.
[c] $J_{SC}$ values integrated from the corresponding $EQE_{LSC}$ spectra were used for *PCE* and *LUE* calculation.
[d] $V_{OC}$ of 1.00 V is assumed for *PCE* and *LUE* calculation comparison, consistent with the range of $V_{OC}$s experimentally measured (1.00-1.02 V)
[e] *FF* of 80% is assumed for *PCE* and *LUE* calculation and comparison, consistent with the range of *FF*s experimentally measured (79-81%).

**Table S2.** An overview of literature reports for LSC/TLSC devices.

| References | Luminophore(s) | *QY%* | Size (cm$^2$) | *G* | *AVT%* | *CRI* | *PCE%* | *EQE$_{LSC}$* | *LUE* |
|---|---|---|---|---|---|---|---|---|---|
| This Work | $Cs_2Mo_6I_8(CF_3CF_2COO)_6$ NCs | 80±5 | 5.08×5.08 | 2 | 75.8 | 88.3 | 3.11 | Yes | 2.36 |
| | BODIPY | 40±3 | | | | | | | |
| 4 | $(TBA)_2Mo_6Cl_{14}$ NCs | 50-55 | 2.5×2.5 | 6.25 | 84.0 | 94.0 | 0.44 | Yes | 0.37 |
| 5 | Cy7-NHS | 20±1 | 2.0×2.0 | 5 | 86.0 | 94.0 | 0.40 | Yes | 0.34 |
| 6 | Cy7-$NEt_2$-I | 26±1 | 5.08×5.08 | 2 | 77.1 | 75.6 | 0.36 | Yes | 0.28 |
| 7 | $CO_i$8DFIC | 25±3 | 5.08×5.08 | 2 | 74.4 | 80.0 | 1.24 | Yes | 0.92 |
| 8 | CdSe/$Cd_{1-x}Zn_xS$ | ~70 | 20.32×20.32 | 31 | 84.8 | 91.0 | 0.525[d] | N/A | 0.45 |
| 9 | Si QDs | 46±5 | 12×12 | 11.54 | 73.0[a] | 84.1 | 0.79[d] | N/A | 0.58 |
| 10 | CdSe/CdS | 45 | 21.5×1.35 | 1.23 | 84.9[a] | 89.2 | N/A[e] | N/A | N/A |
| 11 | SINc:t-U(5000) | 16<br>(UV)<br>8<br>(NIR) | 7.6×2.6 | 9.69 | 89.0[a] | 97.7 | 0.414[d] | Yes | 0.37 |
| 12 | $CuInS_2$/ZnS | 66 | 10×10 | 17.85 | 37.7[a] | 76.9 | 2.18[f] | Yes | 0.82 |
| 13 | Mn:$Cd_xZn_{1-x}S$/ZnS (Top) | 78±2 | 15.24×15.24 | 23.23 | 88.8[b] | 95.5 | 1.3[g] | N/A | N/A |
| | $CuInSe_2$/ZnS (Bottom) | 65-75 | | 23.23 | 8.5[b] | 0.42 | 1.8[g] | N/A | N/A |
| 14 | $CuInSe_xS_{2-x}$/ZnS | 40±4 | 12×12 | 10 | 45.6 | 77.1 | 0.93[d] | N/A | 0.412 |
| 15 | $CuInS_2$/CdS NCs | ~45 | 7.5×7.5 | 6.7 | 60.1[a] | 82.2 | 1.57[d] | N/A | 0.95 |
| 16 | PbS/CdS | 40-50 | 2.0×1.5 | 2.14 | 43.0[a] | 65.6 | 1.68[d] | N/A | 0.72 |
| 17 | bPDI-3 LR 305 | 97.7 | 20×20 | 50 | 46.0[a] | 57.0 | 1.90[d] | N/A | 0.87 |
| 18 | LR 305 LO 240 | ~95 | 3.5×10 | 1.30 | 21.0[a] | 19.0 | 0.23 | N/A | 0.05 |
| 19 | BODIPY Derivatives | 64±1 | 10×10 | 6.25 | 14.0 | N/A | 1.63[g] | N/A | 0.23 |
| | | | | | 53.5 | 75.3 | 1.31[g] | | 0.70 |
| 20 | Zn Al co-doped $CuInS_2$ | N/A | 1.8×1.8 | 4.1 | 82.5 | 99.1 | N/A[h] | N/A | N/A |
| 21 | N-doped Carbon Dots | N/A | 2.0×2.0 | 2.5 | 78.4 | 93.5 | N/A[h] | N/A | N/A |
| 22 | N-doped Carbon Dots | N/A | 2.5×1.6 | 3.03 | 77.7 | 95.6 | N/A[h] | N/A | N/A |
| 23 | BPEA Down-conversion | 85 | 5.0×1.0 | 4.17 | 82.3[a] | 50.3 | N/A[h] | N/A | N/A |
| | BPEA Down-conversion<br>PdTPBP Up-Conversion | 85<br>4 | 5.0×1.0 | 4.17 | 68.7[a] | 42.5 | N/A[h] | N/A | N/A |
| 24 | CdSe@ZnS/ZnS QDs | 79-83<br>(solution) | 10.0×9.0 | 7.9 | 84.4 | 89.7 | 0.337[d] | N/A | 0.284 |

**Table S2.** An overview of literature reports for LSC/TLSC devices (continued).

| References | Luminophore(s) | *QY%* | Size ($cm^2$) | *G* | *AVT%* | *CRI* | *PCE%* | $EQE_{LSC}$ | *LUE* |
|---|---|---|---|---|---|---|---|---|---|
| [25] | LR 305 CRS 040 | N/A | 5.0×5.0 | 2.5 | 0[c] | 0 | 7.1 | Yes | 0 |

[a] Transmission spectrum was acquired with a reference on the reference side of the double-beam spectrometer, so that there is an 8-10% absolute overestimation in *AVT*. These AVT values have been corrected accordingly.
[b] Tandem LSC consists of top and bottom sub-LSCs, however, the total transmission spectrum is not provided.
[c] Reflector placed behind the test LSC as the backdrop therefore, the *AVT* and *CRI* are 0.
[d] Optical efficiencies ($\eta_{OPT}$) were provided. $\eta_{OPT}$ is defined as the ratio of the number of emitted photons reaching the waveguide edge to the number of photon incident on the waveguide front surface over the entire solar spectrum. Therefore, the *PCE* of the LSC device is estimated as: $PCE = \eta_{OPT} \times \eta^{*}_{PV}$, where $\eta^{*}_{PV}$ is the efficiency of edge-mounted PV cell under the waveguided and downshifted flux of the luminophore. $\eta^{*}_{PV}$ is estimated to be 27.6% assuming the highest commercially available Si PV with 22.5% efficiency illuminated under AM 1.5.[26,27]
[e] Neither *PCE* nor $\eta_{OPT}$ were provided.
[f] This *PCE* value is certified.
[g] Area of the edge-mounted PV was used instead of the area of the front surface of the waveguide in *PCE* calculation.
[h] Although *PCE* values calculated from *J-V* characteristics were given, the reported $J_{SC}$ values are above the theoretical SQ limits given the bandgaps (even if their $EQE_{LSC}$ of the corresponding absorption range is 100%, which is impossible given the lower quantum yield and waveguiding losses). Reported data is overestimated by 4-10 ×, due to dividing the $I_{sc}$ by the PV area and not the LSC active area.

## Note 7 Optical Simulation for TLSC Scalability

Optical simulation is provided in Figure S7 and S8 for all the luminophores applied in the dual-band TLSCs. The practical size for the UV component with NC is over 1 m, which is ready for practical deployment. The NIR contribution can be balanced by reducing the concentration (balancing absorption and reabsorption) or increasing Stokes shift (*SS*, or spectral overlap).

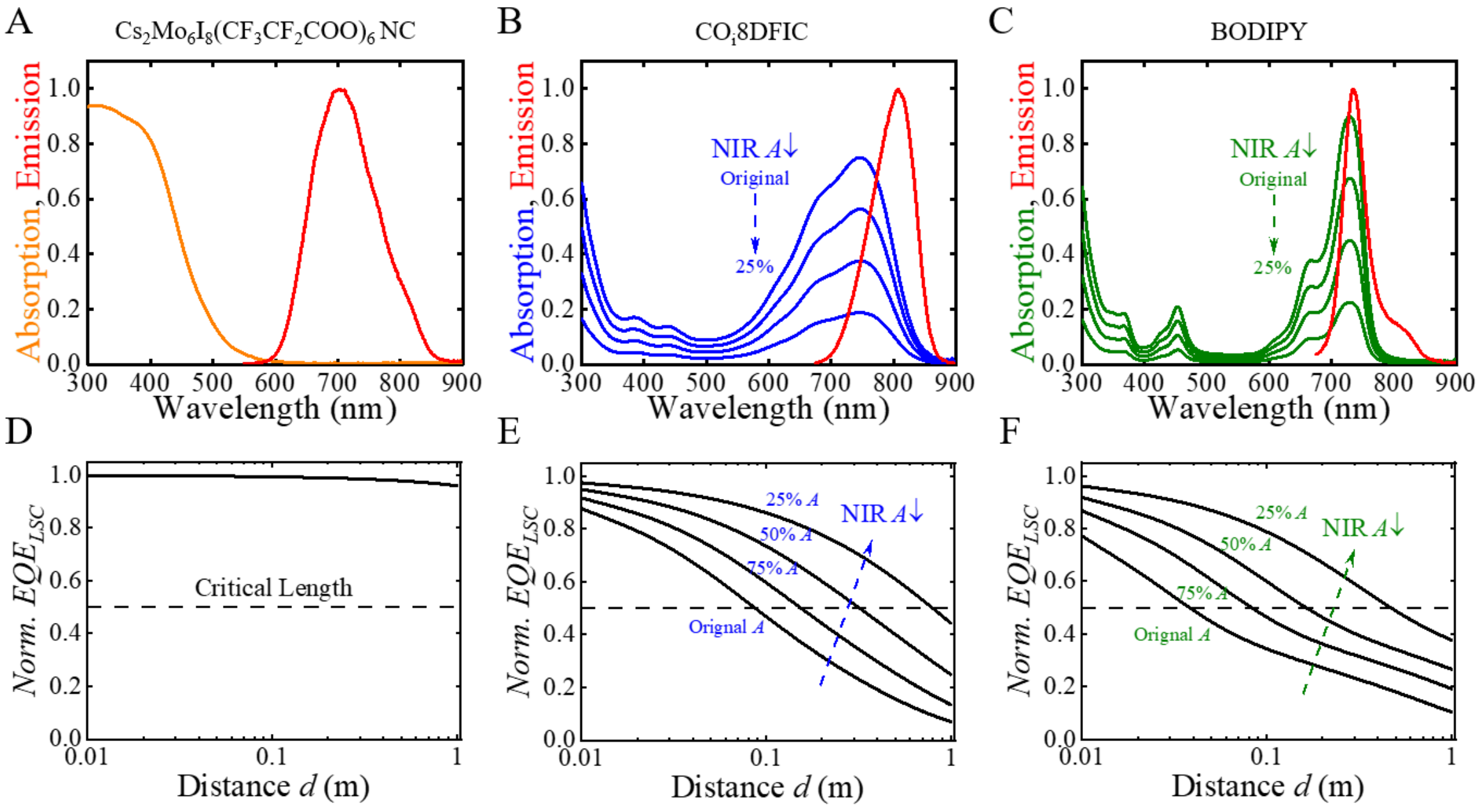

Figure S6. Absolute absorption and normalized emission profiles of (A) NC in UV component TLSC, (B) $CO_i8DFIC$ in NIR component TLSCs with various concentrations and (C) BODIPY in NIR component TLSCs with various concentrations, respectively. (D) to (F) the corresponding normalized $EQE_{LSC}$ as a function of plate length for (A) to (C).

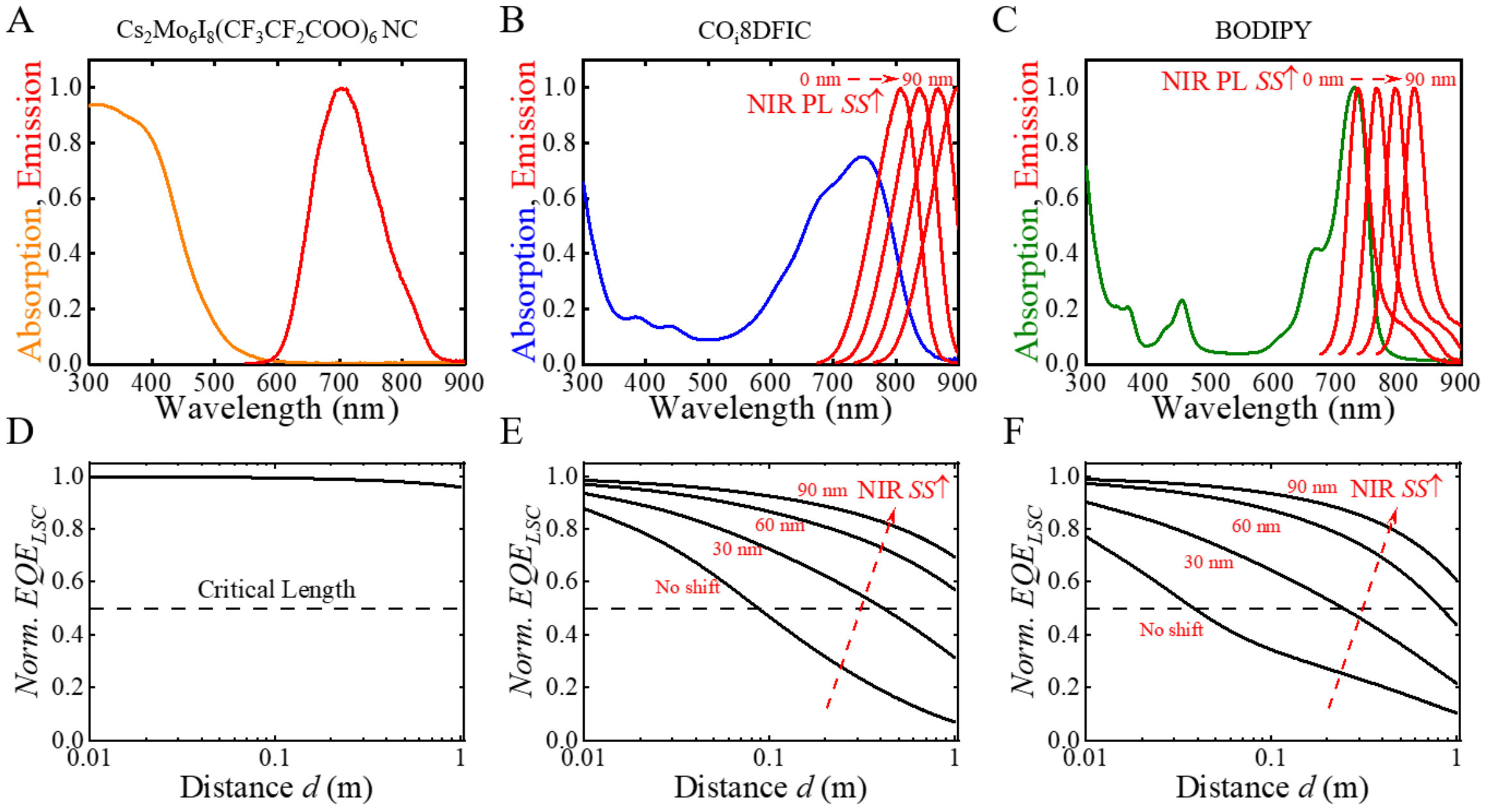

Figure S7. Absolute absorption and normalized emission profiles of (A) NC in UV component TLSC, (B) $CO_i8DFIC$ in NIR component TLSCs with various *SS*s and (C) BODIPY in NIR component TLSCs with various *SS*s, respectively. (D) to (F) the corresponding normalized $EQE_{LSC}$ as a function of plate length for (A) to (C).

## Note 8 CO$_i$8DFIC and BODIPY Synthesis

All chemicals and solvents were purchased from Sigma-Aldrich and used without further purification. Flash chromatography was performed with Silicycle silica gel (60Å, 35-75 μm). Pre-coated 0.25 mm thick silica gel 60 F254 plates (Analtech) were used for analytical TLC and visualized using UV light. NMR ($^{1}$H, $^{13}$C, and $^{19}$F) spectra were recorded with Agilent DirectDrive2 500 MHz spectrometer and referenced with the residual $^{1}$H peak from the deuterated solvents.

**CO$_i$8DFIC**:

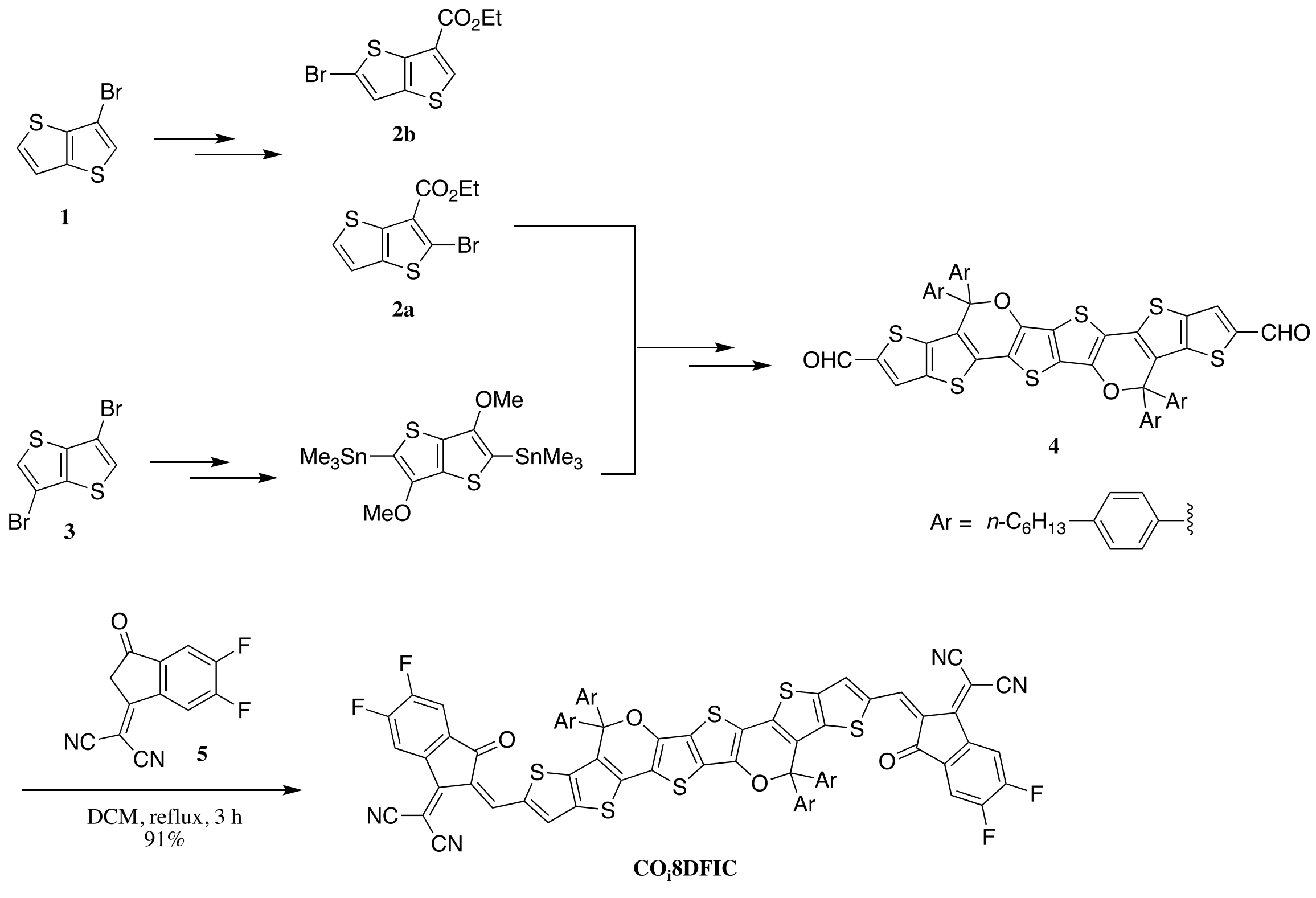

**Figure S8.** General synthesis of **CO$_i$8DFIC**.

Compound **1, 3,** and **5** were purchased from ChemShuttle Co. The synthesis of **CO$_i$8DFIC** closely followed the literature reported procedures,[28,29] with small alterations to a few steps. Briefly, the synthesis commenced with bromide **1**, and in two steps the desired bromo ester **2a** was obtained together with its regioisomer **2b** in a ~2:1 ratio. As reported previously,[7] we used recrystallization instead of column chromatography to separate **2a** from **2b**. Recrystallization of the **2a**/**2b** mixture was done by dissolving the material in pure DCM at room temperature, followed by slow addition of $Et_2O$ to a final 3:1 ratio in volume. The crystals formed after overnight standing at room temperature was collected and washed with cold $Et_2O$ to afford **2a** in substantially higher purity (10:1 **2a**:**2b**), and was used in subsequent transformations without further purification. To avoid yield lowering caused by prolonged reaction time, the final step reaction between **4** and **5** was kept to no longer than 3 h and afforded **COi8DFIC** in 91% yield.

**BODIPY**:

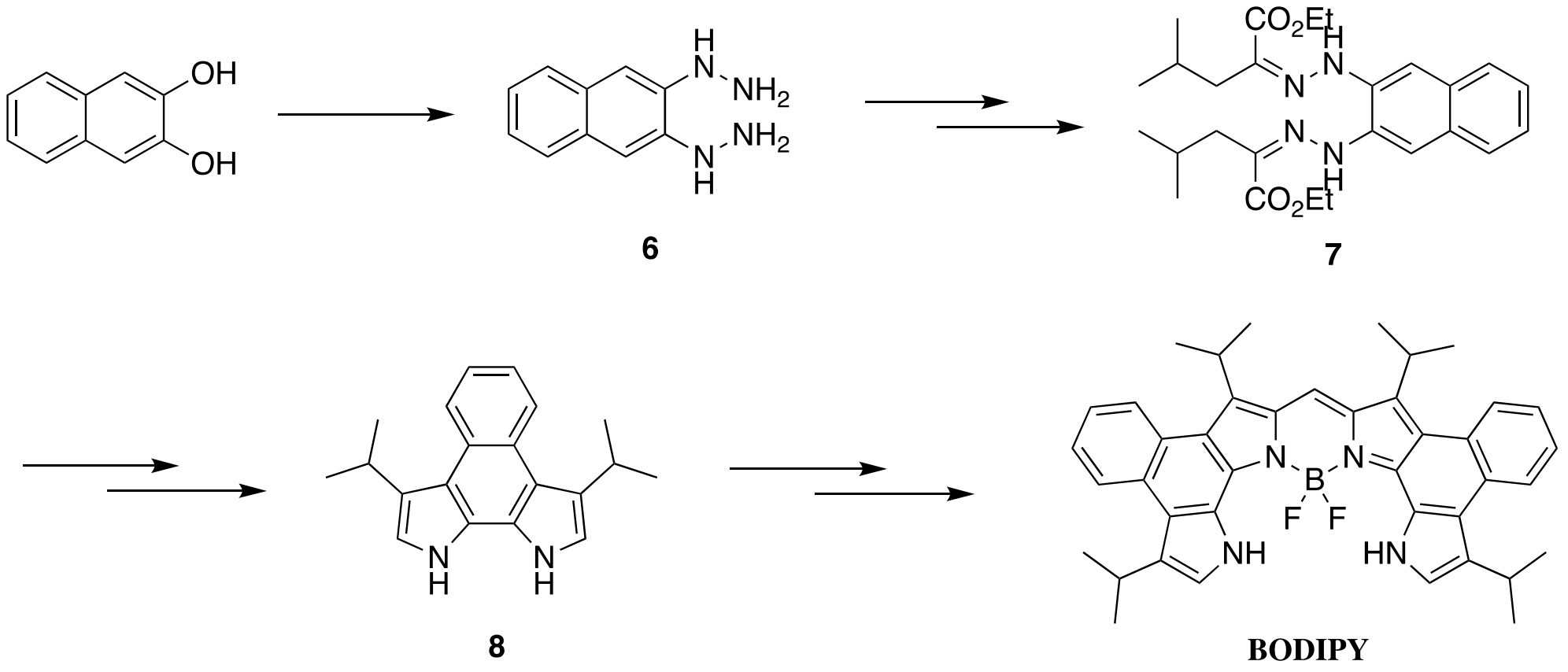

**Figure S9.** General synthesis of **BODIPY**.

The synthesis of BODIPY luminophore followed previously reported procedures with improved yields.[30,31] The synthesis commenced with 2,3-dihydroxynaphthalene, which was converted to 2,3-dihydrazine **6** by treating with hydrazine sulfate and hydrazine hydrate,[32] and subsequently to corresponding dihydrazone with isobutyl ethyl oxalate. Acid-catalyzed Fischer indole synthesis and subsequent decaboxylation furnished key precursor naphthobipyrrole **8**, which was further transformed to target **BODIPY** by reacting with triethyl orthoformate in the presence of $POCl_3$ then treatment with $BF_3{\cdot}OEt_2$.

**Reference**

1. Saito, N., Cordier, S., Lemoine, P., Ohsawa, T., Wada, Y., Grasset, F., Cross, J.S., and Ohashi, N. (2017). Lattice and Valence Electronic Structures of Crystalline Octahedral Molybdenum Halide Clusters-Based Compounds, Cs2[Mo6X14] (X = Cl, Br, I), Studied by Density Functional Theory Calculations. Inorg. Chem. *56*, 6234–6243.

2. Kirakci, K., Kubát, P., Dušek, M., Fejfarová, K., Šícha, V., Mosinger, J., and Lang, K. (2012). A Highly Luminescent Hexanuclear Molybdenum Cluster – A Promising Candidate toward Photoactive Materials. Eur. J. Inorg. Chem. *2012*, 3107–3111.

3. Riehl, L., Ströbele, M., Enseling, D., Jüstel, T., and Meyer, H.-J. (2016). Molecular Oxygen Modulated Luminescence of an Octahedro-hexamolybdenum Iodide Cluster having Six Apical Thiocyanate Ligands. Zeitschrift für Anorg. und Allg. Chemie *642*, 403–408.

4. Zhao, Y., and Lunt, R.R. (2013). Transparent Luminescent Solar Concentrators for Large-Area Solar Windows Enabled by Massive Stokes-Shift Nanocluster Phosphors. Adv. Energy Mater. *3*, 1143–1148.

5. Zhao, Y., Meek, G.A., Levine, B.G., and Lunt, R.R. (2014). Near-infrared harvesting transparent luminescent solar concentrators. Adv. Opt. Mater. *2*, 606–611.

6. Yang, C., Zhang, J., Peng, W.T., Sheng, W., Liu, D., Kuttipillai, P.S., Young, M.,

Donahue, M.R., Levine, B.G., Borhan, B., et al. (2018). Impact of Stokes Shift on the Performance of Near-Infrared Harvesting Transparent Luminescent Solar Concentrators. Sci. Rep. *8*, 16359.

7. Yang, C., Moemeni, M., Bates, M., Sheng, W., Borhan, B., and Lunt, R.R. (2020). High-Performance Near-Infrared Harvesting Transparent Luminescent Solar Concentrators. Adv. Opt. Mater. *n/a*, 1901536.

8. Li, H., Wu, K., Lim, J., Song, H.-J., and Klimov, V.I. (2016). Doctor-blade deposition of quantum dots onto standard window glass for low-loss large-area luminescent solar concentrators. Nat. Energy *1*, 16157.

9. Meinardi, F., Ehrenberg, S., Dhamo, L., Carulli, F., Mauri, M., Bruni, F., Simonutti, R., Kortshagen, U., and Brovelli, S. (2017). Highly efficient luminescent solar concentrators based on earth-abundant indirect-bandgap silicon quantum dots. Nat. Photonics *11*, 177.

10. Meinardi, F., Colombo, A., Velizhanin, K.A., Simonutti, R., Lorenzon, M., Beverina, L., Viswanatha, R., Klimov, V.I., and Brovelli, S. (2014). Large-area luminescent solar concentrators based on /`Stokes-shift-engineered/’ nanocrystals in a mass-polymerized PMMA matrix. Nat Phot. *8*, 392–399.

11. Rondão, R., Frias, A.R., Correia, S.F.H., Fu, L., de Zea Bermudez, V., André, P.S., Ferreira, R.A.S., and Carlos, L.D. (2017). High-Performance Near-Infrared Luminescent Solar Concentrators. ACS Appl. Mater. Interfaces *9*, 12540–12546.

12. Bergren, M.R., Makarov, N.S., Ramasamy, K., Jackson, A., Guglielmetti, R., and McDaniel, H. (2018). High-Performance CuInS2 Quantum Dot Laminated Glass Luminescent Solar Concentrators for Windows. ACS Energy Lett. *3*, 520–525.

13. Wu, K., Li, H., and Klimov, V.I. (2018). Tandem luminescent solar concentrators based on engineered quantum dots. Nat. Photonics *12*, 105–110.

14. Meinardi, F., McDaniel, H., Carulli, F., Colombo, A., Velizhanin, K.A., Makarov, N.S., Simonutti, R., Klimov, V.I., and Brovelli, S. (2015). Highly efficient large-area colourless luminescent solar concentrators using heavy-metal-free colloidal quantum dots. Nat. Nanotechnol. *10*, 878–885.

15. Sumner, R., Eiselt, S., Kilburn, T.B., Erickson, C., Carlson, B., Gamelin, D.R., McDowall, S., and Patrick, D.L. (2017). Analysis of Optical Losses in High-Efficiency CuInS2-Based Nanocrystal Luminescent Solar Concentrators: Balancing Absorption versus Scattering. J. Phys. Chem. C *121*, 3252–3260.

16. Zhou, Y., Benetti, D., Fan, Z., Zhao, H., Ma, D., Govorov, A.O., Vomiero, A., and Rosei, F. (2016). Near Infrared, Highly Efficient Luminescent Solar Concentrators. Adv. Energy Mater. *6*, 1501913-n/a.

17. Zhang, B., Zhao, P., Wilson, L.J., Subbiah, J., Yang, H., Mulvaney, P., Jones, D.J., Ghiggino, K.P., and Wong, W.W.H. (2019). High-Performance Large-Area Luminescence Solar Concentrator Incorporating a Donor–Emitter Fluorophore System. ACS Energy Lett., 1839–1844.

18. Krumer, Z., van Sark, W.G.J.H.M., Schropp, R.E.I., and de Mello Donegá, C. (2017). Compensation of self-absorption losses in luminescent solar concentrators by increasing luminophore concentration. Sol. Energy Mater. Sol. Cells *167*, 133–139.

19. Brzeczek-Szafran, A., Richards, C.J., Lopez, V.M., Wagner, P., and Nattestad, A. (2018). Aesthetically Pleasing, Visible Light Transmissive, Luminescent Solar Concentrators Using a BODIPY Derivative. Phys. status solidi *215*, 1800551.

20. Zhu, M., Li, Y., Tian, S., Xie, Y., Zhao, X., and Gong, X. (2019). Deep-red emitting zinc and aluminium co-doped copper indium sulfide quantum dots for luminescent solar concentrators. J. Colloid Interface Sci. *534*, 509–517.

21. Gong, X., Ma, W., Li, Y., Zhong, L., Li, W., and Zhao, X. (2018). Fabrication of high-performance luminescent solar concentrators using N-doped carbon dots/PMMA mixed matrix slab. Org. Electron. *63*, 237–243.

22. Li, Y., Miao, P., Zhou, W., Gong, X., and Zhao, X. (2017). N-doped carbon-dots for luminescent solar concentrators. J. Mater. Chem. A *5*, 21452–21459.

23. Ha, S.-J., Kang, J.-H., Choi, D.H., Nam, S.K., Reichmanis, E., and Moon, J.H. (2018). Upconversion-Assisted Dual-Band Luminescent Solar Concentrator Coupled for High Power Conversion Efficiency Photovoltaic Systems. ACS Photonics *5*, 3621–3627.

24. Brennan, L.J., Purcell-Milton, F., McKenna, B., Watson, T.M., Gun’ko, Y.K., and

Evans, R.C. (2018). Large area quantum dot luminescent solar concentrators for use with dye-sensitised solar cells. J. Mater. Chem. A *6*, 2671–2680.

25. Slooff, L.H., Bende, E.E., Burgers, A.R., Budel, T., Pravettoni, M., Kenny, R.P., Dunlop, E.D., and Büchtemann, A. (2008). A Luminescent Solar Concentrator with 7.1% power conversion efficiency. Phys. Status Solidi - Rapid Res. Lett. *2*, 257–259.

26. Yang, C., Liu, D., and Lunt, R.R. (2019). How to Accurately Report Transparent Luminescent Solar Concentrators. Joule *3*, 2871–2876.

27. Yang, C., and Lunt, R.R. (2017). Limits of Visibly Transparent Luminescent Solar Concentrators. Adv. Opt. Mater. *5*, 1600851-n/a.

28. Wang, J., Zhang, J., Xiao, Y., Xiao, T., Zhu, R., Yan, C., Fu, Y., Lu, G., Lu, X., Marder, S.R., et al. (2018). Effect of Isomerization on High-Performance Nonfullerene Electron Acceptors. J. Am. Chem. Soc. *140*, 9140–9147.

29. Xiao, Z., Jia, X., Li, D., Wang, S., Geng, X., Liu, F., Chen, J., Yang, S., Russell, T.P., and Ding, L. (2017). 26 mA cm−2 Jsc from organic solar cells with a low-bandgap nonfullerene acceptor. Sci. Bull. *62*, 1494–1496.

30. Sarma, T., Panda, P.K., Anusha, P.T., and Rao, S.V. (2011). Dinaphthoporphycenes: Synthesis and Nonlinear Optical Studies. Org. Lett. *13*, 188–191.

31. Roznyatovskiy, V., Lynch, V., and Sessler, J.L. (2010). Dinaphthoporphycenes. Org. Lett. *12*, 4424–4427.

32. Auwers, K., and Markovits, T. V (1905). Ueber vic. m-Xylenol und ein Tetramethyldiphenochinon. Berichte der Dtsch. Chem. Gesellschaft *38*, 226–237.